\documentclass[twocolumn,times]{aastex63}
\usepackage{amsmath}
\usepackage{newtxmath}
\usepackage{mleftright}
\usepackage{afterpage}
\mleftright

\newcommand{\appropto}{\mathrel{\vcenter{
  \offinterlineskip\halign{\hfil$##$\cr
    \propto\cr\noalign{\kern2pt}\sim\cr\noalign{\kern-2pt}}}}}
\shorttitle{Effect of Differential Rotation on Magnetic Braking of Low-Mass and Solar-Like Stars}
\shortauthors{Ireland et al.}
\begin{document}

\title{Effect of Differential Rotation on Magnetic Braking of Low-Mass and Solar-Like Stars: A Proof-of-Concept Study}

\author[0000-0002-8833-1204]{Lewis G. Ireland}
\affil{Department of Physics and Astronomy, University of Exeter, Stocker Road, Exeter, EX4 4QL, UK}

\author[0000-0001-9590-2274]{Sean P. Matt}
\affil{Department of Physics and Astronomy, University of Exeter, Stocker Road, Exeter, EX4 4QL, UK}

\author[0000-0001-6734-0104]{Charlie R. Davey}
\affil{Department of Physics and Astronomy, University of Exeter, Stocker Road, Exeter, EX4 4QL, UK}

\author[0000-0002-4198-2499]{Owain L. Harris}
\affil{Department of Physics and Astronomy, University of Exeter, Stocker Road, Exeter, EX4 4QL, UK}

\author[0000-0002-4993-0757]{Tobias W. Slade-Harajda}
\affil{Department of Physics and Astronomy, University of Exeter, Stocker Road, Exeter, EX4 4QL, UK}

\author[0000-0002-3020-9409]{Adam J. Finley}
\affil{Department of Astrophysics-AIM, University of Paris-Saclay and University of Paris, CEA, CNRS, Gif-sur-Yvette Cedex 91191, France}

\author[0000-0003-0204-8190]{Claudio Zanni}
\affil{INAF - Osservatorio Astrofisico di Torino, Strada Osservatorio 20, {10025 Pino Torinese}, Italy}

\correspondingauthor{Lewis G. Ireland}
\email{L.G.Ireland@exeter.ac.uk}



\begin{abstract}

% This example manuscript is intended to serve as a tutorial and template for
% authors to use when writing their own AAS Journal articles. The manuscript
% includes a history of \aastex\ and documents the new features in the
% previous version, 6.0, as well as the new features in version 6.1. This
% manuscript includes many figure and table examples to illustrate these new
% features.  Information on features not explicitly mentioned in the article
% can be viewed in the manuscript comments or more extensive online
% documentation. Authors are welcome replace the text, tables, figures, and
% bibliography with their own and submit the resulting manuscript to the AAS
% Journals peer review system.  The first lesson in the tutorial is to remind
% authors that the AAS Journals, the Astrophysical Journal (ApJ), the
% Astrophysical Journal Letters (ApJL), and Astronomical Journal (AJ), all
% have a 250 word limit for the abstract.  If you exceed this length the
% Editorial office will ask you to shorten it.

On the main sequence, low-mass and solar-like stars are observed to spin-down over time, and magnetized stellar winds are thought to be predominantly responsible for this significant angular momentum loss. Previous studies have demonstrated that the wind torque can be predicted via formulations dependent on stellar properties, such as magnetic field strength and geometry, stellar radius and mass, wind mass-loss rate, and stellar rotation rate. Although these stars are observed to experience surface differential rotation, torque formulations so far have assumed solid-body rotation.  Surface differential rotation is expected to affect the rotation of the wind and thus the angular momentum loss. To investigate how differential rotation affects the torque, we use the PLUTO code to perform 2.5D magnetohydrodynamic, axisymmetric simulations of stellar winds, using a colatitude-dependent surface differential rotation profile that is solar-like (i.e., rotation is slower at the poles than the equator). We demonstrate that the torque is determined by the average rotation rate in the wind, so that the net torque is less than that predicted by assuming solid-body rotation at the equatorial rate. The magnitude of the effect is essentially proportional to the magnitude of the surface differential rotation, for example, resulting in a torque for the Sun that is $\sim 20 \%$ smaller than predicted by the solid-body assumption. We derive and fit a semi-analytic formulation that predicts the torque as a function of the equatorial spin rate, magnitude of differential rotation, and wind magnetization (depending on the dipolar magnetic field strength and mass-loss rate, combined).

\end{abstract}

% \keywords{convection --- magnetohydrodynamics (MHD) --- stars: fundamental parameters --- 
% stars: low-mass --- stars: interiors --- stars: rotation}



\section{Introduction}\label{sec:intro}

The evolution of rotation rates of low-mass and solar-like stars on the main sequence (MS) is a complex function of their mass and age, as illustrated, for example, by rotation-period-mass diagrams from observations of large stellar clusters \citep[see, e.g.,][]{Barnes_2003,2009IAUS..258..363I,Meibom_2011,2014prpl.conf..433B,Stauffer_2016,Davenport_2017}. 
Generally, these stars are observed to spin-down with time \citep{Skumanich1972:aa}, due to the extraction of angular momentum by magnetized stellar winds \citep{1958ApJ...128..664P, 1962AnAp...25...18S, 1967ApJ...148..217W, 1968MNRAS.138..359M}.
Models for this rotational evolution \citep[e.g.,][]{1988ApJ...333..236K, 1991ApJ...376..204M,1997A&A...326.1023B, Bouvier_2008, Denissenkov_2010, 2012ApJ...754L..26M, Reiners_2012, van_Saders_2013, Gallet_2013, Gallet_2015, Lanzafame_2015,Matt_2015, 10.1093/mnras/stw369,Gondoin_2017,Garraffo_2018,Amard_2019,Breimann_2021} rely on theoretical formulations for stellar wind torques, which depend on a variety of stellar properties, such as masses and radii, magnetic field strengths and topology, coronal temperatures, and mass-loss rates. 
Due to the importance of magnetized stellar winds for understanding MS rotational evolution, there is a need 
for the development of formulations that predict the torque and include all of the important physical processes in stellar winds.

Early and analytic models of stellar winds and angular momentum loss \citep[e.g.,][]{1958ApJ...128..664P, 1962AnAp...25...18S, 1965SSRv....4..666P, 1967ApJ...148..217W, 1968MNRAS.138..359M, 1988ApJ...333..236K,Sauty_2011} studied thermal-pressure driven winds, in the presence of magnetic fields, demonstrating that initially sub-sonic stellar winds accelerate beyond the sound speed and beyond the magnetic Alfv\'en wave speed (at the Alfv\'en radius).  The angular momentum loss is enhanced by the presence of a magnetic field, and magnetocentrifugal processes in rapid rotators can even enhance the overall wind speed.
\citet{1967ApJ...148..217W} also demonstrated that the stellar wind torque depends on the square of the Alfv\'en radius, which by mechanical analogy is often referred to as the ``magnetic lever arm" length. 
More recently, stellar wind dynamics have been studied numerically via magnetohydrodynamic (MHD) simulations, allowing for the comparison between self-consistently computed values of stellar wind torque and those predicted via parameterizations of the Alfv\'en radius. 
\citet{2008ApJ...678.1109M} investigated how the torque depends on magnetic dipole field strength and mass-loss rate around slow rotators \citep[see also][]{1993MNRAS.262..936W,Cohen_2014}.
Subsequent works explored how the torques depend on magnetocentrifugal effects around fast rotators \citep{2012ApJ...754L..26M}, magnetic fields with more complex magnetic geometries \citep{2008ApJ...678.1109M,  2015ApJ...798..116R, Garraffo_2016, 2017ApJ...845...46F, 2018ApJ...854...78F}, coronal temperature \citep{2017ApJ...849...83P}, and magnetic cycle variations \citep{2018ApJ...864..125F,2018JPlPh..84e7601P,2019ApJ...876...44F}.
So far, studies that have been used to derive formulations for stellar wind torques have assumed that the stellar surface (or more specifically, the base of the wind-emitting region) rotates as a solid body.

Stars behave as rotating fluids, and so the surface rotation rate of a star is not always uniform, but instead can vary as a function of colatitude \citep[see, e.g.,][]{1999A&A...344..911K,Kuker_2011}. 
Common trackers of surface differential rotation for the Sun include sunspots and Doppler measurements \citep[see, e.g.,][for further discussion and comparison between measurements]{2000SoPh..191...47B}.
Typical measured values show the poles to be spinning $\sim 30 \%$ slower than the equator \citep{1990ApJ...351..309S}. 
For other F, G, and K MS stars, the surface differential rotation has been inferred from, for example, Doppler imaging \citep[see, e.g.,][]{10.1046/j.1365-8711.2002.05120.x}, spectral line Fourier transforms \citep[see, e.g.,][]{2002A&A...384..155R}, and photometric measurements of stellar periods \citep[see, e.g.,][]{Reinhold_2013}. 
A commonly used prescription to describe the surface rotation rate as a function of colatitude, $\theta$, is $\Omega (\theta) = \Omega_{\star,\text{eq}} (1 - \alpha \cos^2{\theta}$), where $\Omega_{\star,\text{eq}}$ is the angular rotation rate at the equator, and the relative differential rotation rate is defined as $\alpha = \Delta \Omega / \Omega_{\star,\text{eq}} \equiv (\Omega_{\star,\text{eq}} - \Omega_{\star,\text{pole}}) / \Omega_{\star,\text{eq}}$, where $\Omega_{\star,\text{pole}}$ is the angular rotation rate at the poles.
Thus, for the Sun, $\alpha \approx 0.3$. This relative differential rotation rate is observed to be approximately correlated with rotation period, with the vast majority of F, G, and K MS stars having $\alpha < 0.3$, and most observed samples of rapidly-rotating stars (that rotate faster, on average, than the Sun) showing $\alpha \rightarrow 0$ \citep[see, e.g.,][]{2002AN....323..336C,Reiners393_2002,10.1111/j.1745-3933.2005.08587.x,Reiners_2006,2007AN....328.1030C,10.1093/mnras/stw1443}.
Most stars appear to be undergoing solar-like rotation, where the equator spins faster than the poles ($\alpha > 0$). However, it is possible that some stars are undergoing anti-solar rotation, where the poles are faster than the equator ($\alpha < 0$), or more complex profiles such as cylindrical-banded rotation, which consists of alternating zonal jets \citep[see, e.g.,][and references therein for discussion of solar-like and anti-solar differential rotation]{2017ApJ...836..192B}.

We expect a priori that differential rotation will influence the rotation of the wind, hence the stellar torque. 
In order to quantify and be able to predict the effects of differential rotation on the torque, we perform 74 2.5D axisymmetric stellar wind simulations, changing $\Omega_{\star,\text{eq}}$, $\alpha$ (restricted to solar-like cases), and the dipolar magnetic field strength.
We show that the stellar wind torque is simply predicted in terms of the average (``effective") rotation rate of the wind material. Since the wind emanates from a region surrounding the poles, the ``effective" rotation rate is slower than that of the stellar equator, and so the net torque is smaller than would be predicted assuming a solid-body rotation at the rate of the equator. We find a formulation for predicting this ``effective" rotation rate (and thus the torque) as a function of the differential rotation rate, and other stellar and wind parameters. {In this study, we purely investigate how differential rotation can modify the rotation rate at the base of the wind, and how this affects global angular momentum loss. We are concerned only about the rotation rate of open magnetic field lines, independent of the effects on the coronal dynamics and small scale magnetic field structure.}

In Section~\ref{sec:num_method}, we describe the numerical setup for our simulations, the initial conditions and final solutions, boundary conditions, normalizations, and the parameter space explored in this study. In Section~\ref{sec:sims}, we list our simulations, present their qualitative behavior, and introduce the global quantities we use, such as the mass flow rate, the torque, and the unsigned magnetic flux. In Section~\ref{sec:torque_formulation}, we introduce a torque formulation for the stellar wind that accounts for the differential rotation profile investigated in this parameter study. In Section~\ref{sec:disc_conc}, we discuss and conclude our findings.

\section{Numerical Method}\label{sec:num_method}

\subsection{Numerical Setup}\label{sec:num_setup}

The simulations in this study are solved numerically governed by the following MHD equations:
\begin{align}\label{eq:MHD}
\begin{gathered}
\frac{\partial \rho}{\partial t} + \boldsymbol{\nabla} \cdot (\rho \boldsymbol{v}) = 0, \\
\frac{\partial \rho \boldsymbol{v}}{\partial t} + \boldsymbol{\nabla} \cdot \left[\rho \boldsymbol{v} \boldsymbol{v} + \left(p + \frac{\boldsymbol{B} \cdot \boldsymbol{B}}{8 \pi}\right) \boldsymbol{I} - \frac{\boldsymbol{B}\boldsymbol{B}}{4\pi} \right] = \rho \boldsymbol{g}, \\
\frac{\partial E}{\partial t} + \boldsymbol{\nabla} \cdot \left[\left(E + p + \frac{\boldsymbol{B} \cdot \boldsymbol{B}}{8 \pi}\right) \boldsymbol{v} - \frac{(\boldsymbol{v} \cdot \boldsymbol{B}) \boldsymbol{B}}{4 \pi} \right] = \rho \boldsymbol{g} \cdot \boldsymbol{v}, \\
\frac{\partial \boldsymbol{B}}{\partial t} + \boldsymbol{\nabla} \times (\boldsymbol{B} \times \boldsymbol{v}) = 0,
\end{gathered}
\end{align}
which represent the mass-continuity, momentum, energy, and magnetic induction equation, respectively. Here, $\rho$ represents the mass density, $\boldsymbol{v}$ the velocity field, $p$ the thermal pressure, $\boldsymbol{B}$ the magnetic field, $\boldsymbol{I}$ the identity matrix, and $\boldsymbol{g} = - (GM_\star/R^2) \boldsymbol{\hat{R}}$ the gravitational acceleration ($G$ is the gravitational constant, $M_\star$ is the stellar mass, and $R$ is the spherical radius). The total energy density is written as 
\begin{equation}\label{eq:energy}
	E = \rho u + \rho \frac{\boldsymbol{v} \cdot \boldsymbol{v}}{2} + \frac{\boldsymbol{B} \cdot \boldsymbol{B}}{8\pi},
\end{equation}
where $u$ is specific internal energy (per unit mass). We use a polytropic wind, with an equation of state of the form $\rho u = p/(\gamma - 1)$, where $\gamma$ is the adiabatic index. We numerically solve Equations~(\ref{eq:MHD})-(\ref{eq:energy}) using the PLUTO code \citep{0067-0049-170-1-228,2012ApJS..198....7M}, using a finite-volume Godunov scheme and a linearized ROE Riemann solver \citep{1981JCoPh..43..357R}. We use the constrained transport method to control the divergence-free constraint $\boldsymbol{\nabla} \cdot \boldsymbol{B} = 0$ \citep[see][]{2005JGRA..11012226T}.  

All simulations are 2.5D (2D computational domain, with 3D vector components), and adopt the ($R$,$\theta$) spherical coordinate system. Axisymmetry is assumed about the stellar rotation axis. We define the cylindrical radius using $r=R \sin{\theta}$, where $R$ is the spherical radius. We use a logarithmic grid for the radial direction, consisting of 256 grid cells and covering [1,60]$R_\star$, where $R_\star$ is the stellar radius. We use a uniform grid for the $\theta$ direction, consisting of 512 grid cells and covering [0,$\pi$].

\subsection{Initial Conditions and Final Solutions}\label{sec:init_cond}

We initialize the computational domain with a stellar corona and a magnetic field configuration. The stellar corona is initialized with a 1D spherically symmetric, isotropic, polytropic Parker wind solution, which is defined by the ratio of the stellar surface sound speed $c_{\text{s},\star}=(\gamma p_\star/\rho_\star)^{1/2}$ to the stellar surface escape velocity $v_\text{esc}=(2GM_\star/R_\star)^{1/2}$, where $p_\star$ and $\rho_\star$ are pressure and density at the stellar surface, respectively. We set $\gamma = 1.05$ (near-isothermal), so the wind is heated during expansion without requiring explicit heating terms in the energy equation from Equation~(\ref{eq:MHD}) \citep[see discussion in][and references therein]{2018ApJ...854...78F}. The magnetic field configuration is initially purely dipolar and aligned with the axis of rotation. The $R$ and $\theta$ components are described by

\begin{equation}\label{eq:Br}
B_R(R,\theta) = B_\star \left(\frac{R_\star}{R}\right)^3 \cos{\theta}
\end{equation}
and

\begin{equation}\label{eq:Btheta}
B_\theta(R,\theta) = \frac{1}{2} B_\star \left(\frac{R_\star}{R}\right)^3 \sin{\theta},
\end{equation}
respectively, where $B_\star$ to represents the polar stellar surface magnetic field strength.

We run each simulation until it reaches a steady-state or, in some cases, an oscillating state \citep[as described in][]{2017ApJ...849...83P}, which we regard as the final wind solution.  
The winds are characterized by a continuous and ubiquitous flow from the lower radial boundary (the ``star''), through the grid, and out of the outer radial boundary, at speeds that exceed all information-carrying-wave speeds.
Thus, the final solutions are insensitive to the initial conditions but are instead essentially entirely determined by the lower radial boundary conditions (and somewhat influenced by the axial boundary, at $\theta$=0 and $\pi$), described in the following section.

\subsection{Boundary Conditions}\label{sec:bound_cond}

The computational domain is enclosed by a boundary filled with ghost cells, allowing for the implementation of boundary conditions. At the inner boundary, the density and thermal pressure profiles from the Parker wind solution are kept fixed. The $R$ magnetic field component is kept fixed to conserve the total stellar flux, but the $\theta/\phi$ magnetic field components are free to change via a linear extrapolation. The poloidal velocity $v_\text{p}$ is imposed to be parallel to the poloidal magnetic field $B_\text{p}$, as well as the continuity of the mass flux per magnetic flux (an axisymmetric MHD invariant)
\begin{equation}\label{eq:massflux}
\kappa = \frac{\rho \boldsymbol{v_\text{p}} \cdot \boldsymbol{B_\text{p}}}{\lvert \boldsymbol{B_\text{p}} \rvert^2}
\end{equation}
along magnetic field lines, which ensures smooth inflow of the stellar wind \citep[see, e.g.,][]{Ustyugova_1999,Keppens_2000}. The stellar rotation at the inner boundary is set via the toroidal velocity,

\begin{equation}\label{eq:rot_inner}
v_\phi = r \Omega_\star(\theta) + v_\text{p} \frac{B_\phi}{B_\text{p}},
\end{equation}
where $\Omega_{\star}(\theta)$ is the stellar rotation rate at a given colatitude $\theta$. By using Equation~(\ref{eq:rot_inner}), we impose the differential rotation to the rate of rotation of the magnetic surfaces. The toroidal speed of the plasma (which determines, e.g., the Doppler shifts) slightly deviates from this, and therefore actually contains some very small degree of differential rotation (even for a solid rotator). On the other hand, since $v_\text{p}$ (the injection speed of the wind) is smaller than $r\Omega_\star$ (at least for faster rotators), and $B_\phi/B_\text{p}$ should be also reasonably small, the two speeds (plasma and field) almost coincide. We adopt the following simple differential rotation profile for the stellar surface rotation rate:
\begin{equation}\label{eq:DR_profile}
\Omega_{\star}(\theta) = \Omega_{\star,\text{eq}} (1 - \alpha \cos^2{\theta}),
\end{equation}
where $\Omega_{\star,\text{eq}}$ is the angular rotation rate at the stellar equator, and $\alpha = \Delta \Omega / \Omega_{\star,\text{eq}} \equiv (\Omega_{\star,\text{eq}} - \Omega_{\star,\text{pole}}) / \Omega_{\star,\text{eq}}$, where $\Omega_{\star,\text{pole}}$ is the angular rotation rate at the poles. This rotation profile is kept fixed throughout computation. At the outer boundary, all quantities have vanishing derivatives, e.g., $d v_R / dR = 0$, allowing them to flow outward from the computational domain.

\subsection{Simulation Parameters}\label{sec:sim_params}

We investigate a parameter space where we systematically vary the following quantities:

\begin{enumerate}
\item{Surface polar magnetic field strength, $B_\star$; in practice, this is controlled via the input parameter $v_\text{A}/v_\text{esc}$, i.e., the ratio of the surface polar Alfv\'en velocity $v_\text{A} = B_\star / (4 \pi \rho_\star)^{1/2}$ and the stellar escape velocity $v_\text{esc}=(2GM_\star/R_\star)^{1/2}$.}
\item{Stellar equatorial rotation rate, expressed as a fraction of the break-up rate, $f_\text{eq}=\Omega_{\star,\text{eq}} R_\star / v_{\text{K},\star}$;}
\item{Relative differential rotation rate, $\alpha=(\Omega_{\star,\text{eq}} - \Omega_{\star,\text{pole}})/\Omega_{\star,\text{eq}}$;}
\end{enumerate}
where $\Omega_{\star,\text{eq}}$ and $\Omega_{\star,\text{pole}}$ are the equatorial and polar stellar rotation rates, respectively. We fix $\gamma=1.05$ and the stellar wind sound speed at the stellar surface to $c_{\text{s},\star} = 0.25 v_\text{esc}$, for all simulations.

\subsection{Units and Normalization}\label{sec:units}

{We perform simulations in dimensionless units; here, we list normalization factors required to convert quantities into physical units representative of different types of stars. We express length in units of $R_\star$, density in units of its value at the base of the corona, $\rho_\star$, and velocities in units of stellar surface Keplerian velocity, $v_{\text{K},\star}=(GM_\star/R_\star)^{1/2}$ (where $M_\star$ is the stellar mass). The following normalizations can then be derived from the aforementioned base dimensionalizations: time in units of $t_0 = R_\star/v_{\text{K},\star}$, magnetic field strength in units of $B_0 = (4 \pi \rho_\star v_{\text{K},\star}^2)^{1/2}$, mass flow rate in units of $\dot{M}_0 = \rho_\star R_\star^2 v_{\text{K},\star}$, torque in units of $\dot{J}_0 = \rho_\star R_\star^3 v_{\text{K},\star}^2$, and magnetic flux in units of $\Phi_0 = (4 \pi \rho_\star R_\star^4 v_{\text{K},\star}^2)^{1/2}$.}

{By adopting solar values for radius, $R_\odot=6.96 \times10^{10}$ cm, mass, $M_\odot=1.99 \times10^{33}$ g, and density at the base of the corona, $\rho_\odot = 2.46 \times 10^{-16}$ g cm$^{-3}$ (see Section~\ref{sec:sims}), we can make direct comparisons with main-sequence solar-like and low-mass stars using the following normalizations:}

\begin{align}\label{eq:norm}
\begin{gathered}
v_{\text{K},\star}  = 437 \left(\frac{M_\star}{M_\odot}\right)^{1/2}  \left(\frac{R_\star}{R_\odot}\right)^{-1/2}  \,  \text{km s}^{-1}\\
B_0  = 2.43 \left(\frac{\rho_\star}{\rho_\odot}\right)^{1/2}  \left(\frac{M_\star}{M_\odot}\right)^{1/2}  \left(\frac{R_\star}{R_\odot}\right)^{-1/2} \,  \text{G} \\
t_0 = 0.0184 \left(\frac{M_\star}{M_\odot}\right)^{-1/2} \left(\frac{R_\star}{R_\odot}\right)^{3/2} \, \text{days} \\
\dot{M}_0 =  8.26 \times 10^{-13} \left(\frac{\rho_\star}{\rho_\odot}\right) \left(\frac{M_\star}{M_\odot}\right)^{1/2} \left(\frac{R_\star}{R_\odot}\right)^{3/2} \, M_\odot \, \text{yr}^{-1} \\
\dot{J}_0 =  1.58 \times 10^{32} \left(\frac{\rho_\star}{\rho_\odot}\right) \left(\frac{M_\star}{M_\odot}\right) \left(\frac{R_\star}{R_\odot}\right)^{2} \, \text{erg} \\
\Phi_0 = 1.18 \times 10^{22} \left(\frac{\rho_\star}{\rho_\odot}\right)^{1/2} \left(\frac{M_\star}{M_\odot}\right)^{1/2} \left(\frac{R_\star}{R_\odot}\right)^{3/2} \, \text{Mx}.
\end{gathered}
\end{align}
{Furthermore, assuming the solar fraction of the equatorial break-up rate $f_{\odot,\text{eq}} = 4.73 \times 10^{-3}$ (using a typical measured value of the solar equatorial sidereal rotation period, 24.47 days \citep{1990ApJ...351..309S}), the equatorial stellar rotation period ($P_{\star,\text{eq}} = 2 \pi / \Omega_{\star,\text{eq}} \equiv 2 \pi R_\star / (f_{\text{eq}} v_{\text{K},\star})$) can be expressed as}

\begin{equation}\label{eq:period}
P_{\star,\text{eq}} = 24.47 \left(\frac{f_\text{eq}}{f_{\odot,\text{eq}}}\right)^{-1} \left(\frac{M_\star}{M_\odot}\right)^{-1/2} \left(\frac{R_\star}{R_\odot}\right)^{3/2} \, \text{days}.
\end{equation}

\section{Simulations of Stellar Winds}\label{sec:sims}

This study consists of 74 simulations in total, exploring the parameter space outlined in Section~\ref{sec:sim_params}. Each simulation is ran for a period roughly corresponding to 16 stellar rotation periods for $f_\text{eq} = 0.05$ ($t_0=2000$). Typically, this corresponds to weeks or months in simulated time, where changes in the stellar rotation rate are negligible, justifying our use of fixed $f_\text{eq}$. All simulations evolve to a quasi-steady state in this simulation time period. Simulation input and output parameters can be found in Table~\ref{tab:Pluto_sims}. 

{In Table~\ref{tab:norm_sims}, we make direct comparisons between our simulations and examples of solar-like and low-mass stars, by dimensionalizing our numerical simulation units using normalizations shown in Section~\ref{sec:units}. For $M_\star/M_\odot = (0.10, 0.25, 0.50, 1.00)$, we list range of surface magnetic field strengths and stellar rotation periods that are covered by our parameter space.  For each stellar mass, we also list the ranges mass-loss rate, torque, and open magnetic flux that result from these simulations. Stellar radii for each mass is taken from the grid of low-mass stellar models produced by \citet{Amard_2019} (at 5 Gyr). For the $M_\star = M_\odot$ case, we pick a value of $\rho_\star$ so that the mass-loss rate of model 1 (the closest to a representative solar case) is $\dot{M} = \dot{M}_\odot \approx 1.78 \times 10^{-14}$ $M_\odot$ yr$^{-1}$, which is a reasonable value for the Sun.\footnote{This solar mass-loss rate is the mean of 27 day averages via measurements of the solar wind speed and density from the \emph{ACE} spacecraft \citep[see][]{2018ApJ...864..125F}.}
For $M_\star = 0.1 M_\odot$, we pick a value of $\rho_\star$ so that the range of $B_\star$ in our simulations ranges from $100-800$ G, which is within the range bracketed by Zeeman-Doppler Magnetogram observations \citep[see,e.g.,][]{10.1111/j.1365-2966.2010.17101.x,2019ApJ...876..118S,2019ApJ...886..120S}\footnote{\citet{2019ApJ...886..120S} give a range of dipolar field strengths averaged over the stellar surface, $\langle B_\text{d} \rangle$, therefore values of $B_\star$ are likely to be within a factor of two.}. For the intermediate masses, $M_\star = 0.25 - 0.50$ $M_\odot$, we choose interpolated values of $\rho_\star$ assuming $\rho_\star \appropto M_\star^{-3}$ (given the three orders of magnitude decrease in $\rho_\star$ for one order of magnitude increase in stellar mass), for simplicity.}

\begin{deluxetable*}{ccccc|cccccc||ccccc|cccccc}
\tabletypesize{\footnotesize}

\tablecaption{Variable Input Parameters and Outputted Global Variables for All Simulations \label{tab:Pluto_sims}}

\setlength{\tabcolsep}{2.5pt}
\tablehead{\rule{0pt}{4ex} Model & $f_{\text{eq}}$ & $\alpha$ & $\frac{v_\text{A}}{v_\text{esc}}$ & $\frac{B_\star}{B_0}$$^{\text{a}}$ & $\frac{\dot{M}}{\dot{M}_0}$ & $\frac{\dot{J}}{\dot{J}_0}$ & $\frac{\Phi_\text{open}}{\Phi_0}$ & $\frac{\langle r_\text{A,eq} \rangle}{R_\star}$ & $\Upsilon_\star$ & $\omega_\text{emp}$ & Model & $f_{\text{eq}}$ & $\alpha$ & $\frac{v_\text{A}}{v_\text{esc}}$ & $\frac{B_\star}{B_0}$$^{\text{a}}$ & $\frac{\dot{M}}{\dot{M}_0}$ & $\frac{\dot{J}}{\dot{J}_0}$ & $\frac{\Phi_\text{open}}{\Phi_0}$ & $\frac{\langle r_\text{A,eq} \rangle}{R_\star}$ & $\Upsilon_\star$ & $\omega_\text{emp}$ \\
 &  & & & & $[10^{-2}]$ & $[10^{-2}]$ &  &  & $[10^{4}]$ & & & & & & & $[10^{-2}]$ & $[10^{-2}]$ &  &  & $[10^{4}]$ & }
\startdata
1 & 0.001 & 0 & 1 & 1.41 & 2.15 & 0.102 & 2.87 & 6.89 & 0.26 & 0.946 & 38 & 0.05 & 0.25 & 2 & 2.83 & 2.1 & 8.95 & 5.2 & 9.23 & 1.06 & 0.827 \\
2 & 0.001 & 0 & 2 & 2.83 & 1.67 & 0.184 & 4.44 & 10.5 & 1.34 & 0.943 & 39 & 0.05 & 0.25 & 4 & 5.66 & 1.69 & 17.1 & 8.3 & 14.2 & 5.3 & 0.859 \\
3 & 0.001 & 0 & 4 & 5.66 & 1.21 & 0.321 & 6.44 & 16.3 & 7.41 & 0.946 & 40 & 0.05 & 0.25 & 8 & 11.3 & 1.43 & 32.3 & 13.1 & 21.3 & 25 & 0.865 \\
4 & 0.001 & 0 & 8 & 11.3 & 0.886 & 0.628 & 9.72 & 26.6 & 40.3 & 1.05 & 41 & 0.05 & 0.5 & 1 & 1.41 & 2.37 & 3.36 & 3.07 & 5.33 & 0.236 & 0.598 \\
5 & 0.001 & 0.25 & 1 & 1.41 & 2.15 & 0.0816 & 2.87 & 6.16 & 0.26 & 0.757 & 42 & 0.05 & 0.5 & 2 & 2.83 & 1.95 & 6.36 & 4.97 & 8.07 & 1.14 & 0.607 \\
6 & 0.001 & 0.25 & 2 & 2.83 & 1.67 & 0.145 & 4.44 & 9.32 & 1.34 & 0.744 & 43 & 0.05 & 0.5 & 4 & 5.66 & 1.49 & 11.5 & 7.68 & 12.4 & 5.99 & 0.609 \\
7 & 0.001 & 0.25 & 4 & 5.66 & 1.21 & 0.25 & 6.44 & 14.4 & 7.41 & 0.736 & 44 & 0.05 & 0.5 & 8 & 11.3 & 1.23 & 22.4 & 12.2 & 19.1 & 29.1 & 0.642 \\
8 & 0.001 & 0.25 & 8 & 11.3 & 0.886 & 0.484 & 9.74 & 23.4 & 40.3 & 0.808 & 45 & 0.05 & 0.75 & 1 & 1.41 & 2.27 & 2.2 & 2.98 & 4.41 & 0.246 & 0.399 \\
9 & 0.001 & 0.5 & 1 & 1.41 & 2.15 & 0.0612 & 2.87 & 5.34 & 0.26 & 0.568 & 46 & 0.05 & 0.75 & 2 & 2.83 & 1.82 & 3.84 & 4.74 & 6.49 & 1.23 & 0.378 \\
10 & 0.001 & 0.5 & 2 & 2.83 & 1.67 & 0.106 & 4.44 & 7.97 & 1.34 & 0.544 & 47 & 0.05 & 0.75 & 4 & 5.66 & 1.35 & 6.47 & 7.15 & 9.78 & 6.62 & 0.36 \\
11 & 0.001 & 0.5 & 4 & 5.66 & 1.21 & 0.178 & 6.43 & 12.2 & 7.41 & 0.526 & 48 & 0.05 & 0.75 & 8 & 11.3 & 1.05 & 11.9 & 11 & 15.1 & 34.2 & 0.366 \\
12 & 0.001 & 0.5 & 8 & 11.3 & 0.886 & 0.34 & 9.75 & 19.6 & 40.3 & 0.568 & 49 & 0.1 & 0 & 1 & 1.41 & 4.12 & 14.6 & 4.06 & 5.95 & 0.135 & 1.02 \\
13 & 0.001 & 0.75 & 1 & 1.41 & 2.15 & 0.0409 & 2.87 & 4.36 & 0.26 & 0.38 & 50 & 0.1 & 0 & 2 & 2.83 & 3.75 & 29.3 & 6.95 & 8.84 & 0.596 & 1.05 \\
14 & 0.001 & 0.75 & 2 & 2.83 & 1.67 & 0.0671 & 4.43 & 6.34 & 1.34 & 0.344 & 51 & 0.1 & 0 & 4 & 5.66 & 3.13 & 53.6 & 11.1 & 13.1 & 2.85 & 1.03 \\
15 & 0.001 & 0.75 & 4 & 5.66 & 1.21 & 0.107 & 6.44 & 9.44 & 7.41 & 0.315 & 52 & 0.1 & 0 & 8 & 11.3 & 2.57 & 90.3 & 17 & 18.6 & 13.9 & 0.912 \\
16 & 0.001 & 0.75 & 8 & 11.3 & 0.885 & 0.196 & 9.74 & 14.9 & 40.4 & 0.327 & 53 & 0.1 & 0.25 & 1 & 1.41 & 3.55 & 11.5 & 3.85 & 5.68 & 0.157 & 0.856 \\
17 & 0.01 & 0 & 1 & 1.41 & 2.17 & 1.03 & 2.88 & 6.88 & 0.257 & 0.948 & 54 & 0.1 & 0.25 & 2 & 2.83 & 3.21 & 23 & 6.57 & 8.48 & 0.696 & 0.887 \\
18 & 0.01 & 0 & 2 & 2.83 & 1.71 & 1.91 & 4.53 & 10.6 & 1.3 & 0.969 & 55 & 0.1 & 0.25 & 4 & 5.66 & 2.59 & 40.5 & 10.4 & 12.5 & 3.46 & 0.846 \\
19 & 0.01 & 0 & 4 & 5.66 & 1.24 & 3.31 & 6.58 & 16.4 & 7.22 & 0.961 & 56 & 0.1 & 0.25 & 8 & 11.3 & 2.28 & 74.4 & 16.5 & 18.1 & 15.7 & 0.811 \\
20 & 0.01 & 0 & 8 & 11.3 & 0.941 & 6.67 & 10.2 & 26.6 & 38 & 1.08 & 57 & 0.1 & 0.5 & 1 & 1.41 & 3.04 & 8.23 & 3.59 & 5.2 & 0.184 & 0.656 \\
21 & 0.01 & 0.25 & 1 & 1.41 & 2.16 & 0.821 & 2.88 & 6.16 & 0.258 & 0.76 & 58 & 0.1 & 0.5 & 2 & 2.83 & 2.66 & 16.1 & 6.02 & 7.79 & 0.839 & 0.672 \\
22 & 0.01 & 0.25 & 2 & 2.83 & 1.71 & 1.5 & 4.52 & 9.39 & 1.31 & 0.763 & 59 & 0.1 & 0.5 & 4 & 5.66 & 2.12 & 28.4 & 9.43 & 11.6 & 4.21 & 0.643 \\
23 & 0.01 & 0.25 & 4 & 5.66 & 1.24 & 2.65 & 6.7 & 14.7 & 7.22 & 0.773 & 60 & 0.1 & 0.5 & 8 & 11.3 & 1.82 & 52.6 & 15.2 & 17 & 19.6 & 0.63 \\
24 & 0.01 & 0.25 & 8 & 11.3 & 0.915 & 4.94 & 10 & 23.4 & 39 & 0.82 & 61 & 0.1 & 0.75 & 1 & 1.41 & 2.63 & 5.25 & 3.33 & 4.46 & 0.213 & 0.444 \\
25 & 0.01 & 0.5 & 1 & 1.41 & 2.16 & 0.615 & 2.87 & 5.34 & 0.259 & 0.57 & 62 & 0.1 & 0.75 & 2 & 2.83 & 2.2 & 9.54 & 5.41 & 6.58 & 1.01 & 0.431 \\
26 & 0.01 & 0.5 & 2 & 2.83 & 1.7 & 1.1 & 4.5 & 8.04 & 1.32 & 0.558 & 63 & 0.1 & 0.75 & 4 & 5.66 & 1.73 & 16.9 & 8.53 & 9.87 & 5.15 & 0.419 \\
27 & 0.01 & 0.5 & 4 & 5.66 & 1.23 & 1.89 & 6.67 & 12.4 & 7.27 & 0.551 & 64 & 0.1 & 0.75 & 8 & 11.3 & 1.39 & 28 & 13 & 14.2 & 25.6 & 0.376 \\
28 & 0.01 & 0.5 & 8 & 11.3 & 0.905 & 3.45 & 9.89 & 19.6 & 39.5 & 0.574 & 65 & 0.2 & 0 & 1 & 1.41 & 10.4 & 42.4 & 5.52 & 4.52 & 0.0538 & 1.04 \\
29 & 0.01 & 0.75 & 1 & 1.41 & 2.15 & 0.411 & 2.87 & 4.37 & 0.259 & 0.381 & 66 & 0.2 & 0 & 2 & 2.83 & 9.77 & 85.9 & 9.63 & 6.63 & 0.229 & 1.07 \\
30 & 0.01 & 0.75 & 2 & 2.83 & 1.68 & 0.681 & 4.46 & 6.37 & 1.33 & 0.348 & 67 & 0.2 & 0 & 4 & 5.66 & 7.97 & 147 & 15 & 9.64 & 1.12 & 0.978 \\
31 & 0.01 & 0.75 & 4 & 5.66 & 1.21 & 1.11 & 6.56 & 9.57 & 7.39 & 0.325 & 68 & 0.2 & 0 & 8 & 11.3 & 6.26 & 238 & 22.9 & 13.9 & 5.71 & 0.853 \\
32 & 0.01 & 0.75 & 8 & 11.3 & 0.892 & 1.82 & 9.25 & 14.3 & 40 & 0.302 & 69 & 0.2 & 0.5 & 1 & 1.41 & 6.11 & 24.9 & 4.89 & 4.51 & 0.0914 & 0.748 \\
33 & 0.05 & 0 & 1 & 1.41 & 2.66 & 5.85 & 3.28 & 6.62 & 0.21 & 0.986 & 70 & 0.2 & 0.5 & 2 & 2.83 & 5.5 & 47.7 & 8.29 & 6.59 & 0.406 & 0.739 \\
34 & 0.05 & 0 & 2 & 2.83 & 2.29 & 11.8 & 5.47 & 10.1 & 0.974 & 1.05 & 71 & 0.2 & 0.5 & 4 & 5.66 & 4.2 & 75.9 & 12.7 & 9.53 & 2.13 & 0.649 \\
35 & 0.05 & 0 & 4 & 5.66 & 1.87 & 22.7 & 8.85 & 15.5 & 4.76 & 1.09 & 72 & 0.25 & 0 & 2 & 2.83 & 14.8 & 126 & 10.8 & 5.83 & 0.151 & 1.08 \\
36 & 0.05 & 0 & 8 & 11.3 & 1.61 & 42.6 & 14.1 & 23 & 22.2 & 1.08 & 73 & 0.25 & 0 & 4 & 5.66 & 11.9 & 211 & 16.9 & 8.46 & 0.749 & 0.967 \\
37 & 0.05 & 0.25 & 1 & 1.41 & 2.5 & 4.58 & 3.17 & 6.05 & 0.223 & 0.794 & 74 & 0.3 & 0 & 4 & 5.66 & 17.8 & 299 & 18.7 & 7.54 & 0.502 & 1 \\
\enddata

\tablenotetext{a}{$B_\star$ is not a fundamental input parameter, but it is simply derived from $v_\text{A}/v_\text{esc}$ and tabulated here, for convenience.}

\end{deluxetable*}

\begin{deluxetable*}{cccccc|ccc}
\tabletypesize{\footnotesize}

\tablecaption{Example Ranges of Global Properties in Physical Units (for Our Parameter Space) for $M_\star = 0.1 - 1$ $M_\odot$ (5 Gyr) \label{tab:norm_sims}}
\setlength{\tabcolsep}{3.5pt}
\tablehead{\rule{0pt}{4ex} $M_\star$ & $R_\star$ & $v_{\text{K},\star}$ & $\rho_\star$ & $B_\star$ & $P_{\star,\text{eq}}$ & $\dot{M}$ & $\dot{J}$ & $\Phi_\text{open}$  \\
$[M_\odot]$ & $[R_\odot]$ & $[$km s$^{-1}]$ & $[$g cm$^{-3}]$ & [G] & [days] & $[M_\odot$ yr$^{-1}]$ &  [erg] & [Mx]}
\startdata
0.10 & 0.117 & 404 & $2.46 \times 10^{-13}$ & $100 - 800$ & $0.0489 - 14.7$ & $9.23 \times 10^{-14}$ - $1.85 \times 10^{-12}$ & $8.84 \times 10^{28}$ - $6.46 \times 10^{32}$ & $1.35 \times 10^{22}$ - $1.08 \times 10^{23}$  \\
0.25 & 0.255 & 432 & $1.57 \times 10^{-14}$ & $27.2 - 218$ & $0.0995 - 29.8$& $3.02 \times 10^{-14}$ - $6.06 \times 10^{-13}$ & $6.75 \times 10^{28}$ - $4.93 \times 10^{32}$ & $1.74 \times 10^{22}$ - $1.39 \times 10^{23}$  \\
0.50 & 0.455 & 458 & $1.97 \times 10^{-15}$ & $10.2 - 81.5$ & $0.168 - 50.3$ & $1.27 \times 10^{-14}$ - $2.55 \times 10^{-13}$ & $5.37 \times 10^{28}$ - $3.92 \times 10^{32}$ & $2.07 \times 10^{22}$ - $1.65 \times 10^{23}$  \\
1.00 & 1.00 & 437 & $2.46 \times 10^{-16}$ & $3.44 - 27.5$ & $0.386 - 116$ & $7.32 \times 10^{-15}$ - $1.47 \times 10^{-13}$ & $6.48 \times 10^{28}$ - $4.73 \times 10^{32}$ & $3.37 \times 10^{22}$ - $2.69 \times 10^{23}$  \\
\enddata

\end{deluxetable*}

{There are relatively few observational constraints on the mass loss rate of low-mass and solar-like stars. However, for example, \citet{Wood_2021} show that stars within our mass range in Table~\ref{tab:norm_sims} have mass loss rates of the order of the solar value, with a potential scatter of a few orders of magnitude. We find that the range of $\dot{M}$ shown in Table~\ref{tab:norm_sims} are bracketed by the observational range suggested by \citet{Wood_2021}.
Regarding the rotation rates, we chose a range to cover a parameter space that is well within the slowly-rotating regime, and also pushing into the rapidly-rotating regime.
Consequently, the range of rotation periods in Table~\ref{tab:norm_sims} for the solar case includes the range typically observed \citep[see, e.g.][]{2014prpl.conf..433B}, but it also extends to much longer periods than what is applicable for solar-like stars.
On the other hand, for the low-mass star cases, our parameter space includes the relatively fast-rotation side of the observed rotation rate distributions, but it does not extend to rotational periods as long as observed in the oldest stars \citep{2014prpl.conf..433B}. 
However, stars with longer rotation periods than those covered by our parameter space should still be compatible with our torque formulations because they are in they are in the slowly-rotating regime, where changes in angular momentum are linear with stellar rotation rate.}

% \begin{deluxetable}{cc}
% \tablecaption{Normalization of Each Parameter Type in Table~\ref{tab:Pluto_sims}.\label{tab:conversion}}

% \tablehead{Parameter & Normalization}

% \startdata
% $B_0$ & $(4 \pi \rho_\star v_{\text{K},\star}^2)^{1/2}$ \\
% $\dot{M}_0$ & $\rho_\star R_\star^2 v_{\text{K},\star}$ \\
% $\dot{J}_0$ & $\rho_\star R_\star^3 v_{\text{K},\star}^2$ \\ 
% $\Phi_0$ & $(4 \pi \rho_\star R_\star^4 v_{\text{K},\star}^2)^{1/2}$ \\
% \enddata
% \end{deluxetable}

\subsection{Morphology of Stellar Wind Systems}\label{sec:qualit}

\begin{figure*}
\begin{center}
\includegraphics[width=0.284175\textwidth]{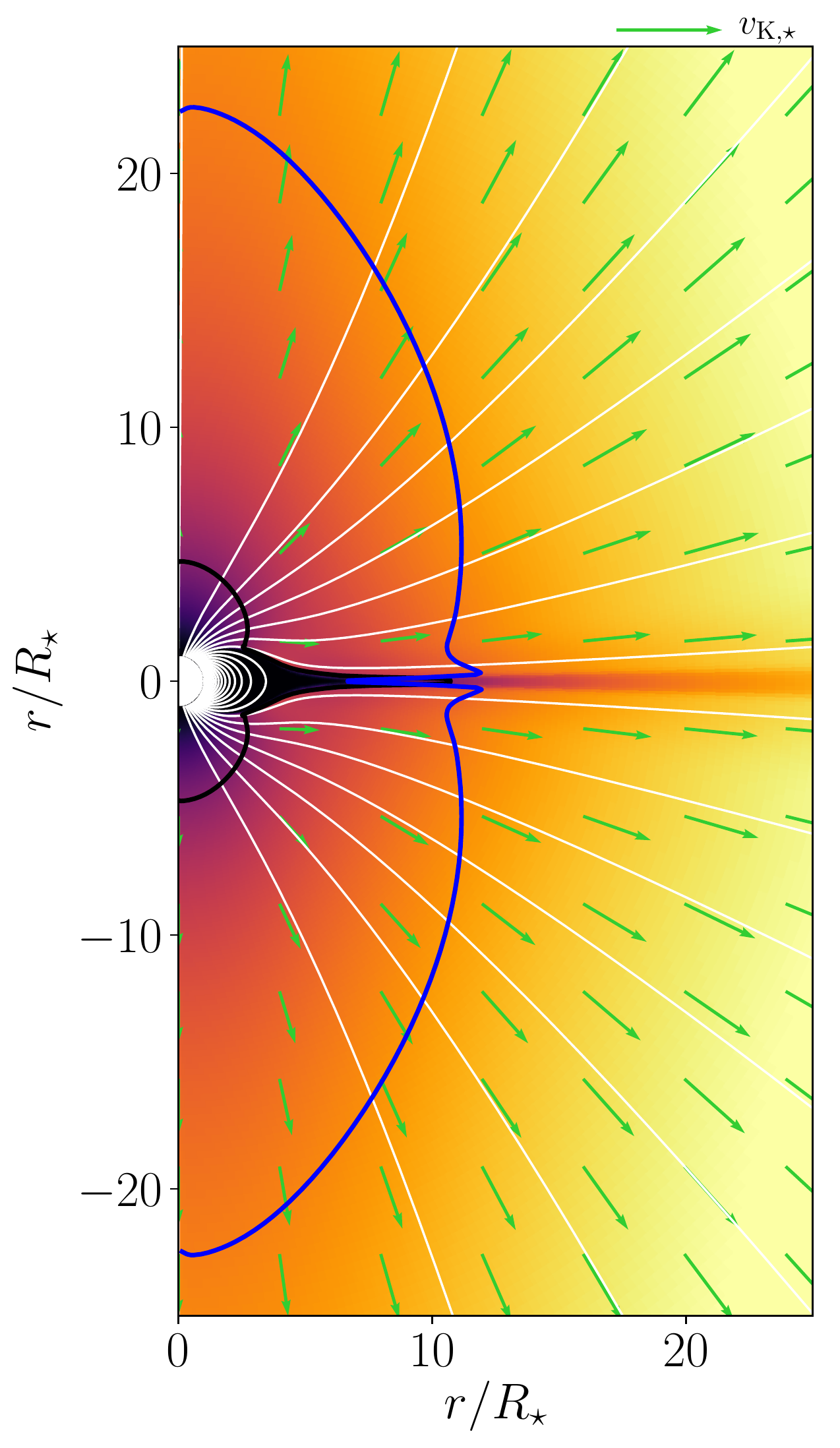}
\includegraphics[width=0.266175\textwidth]{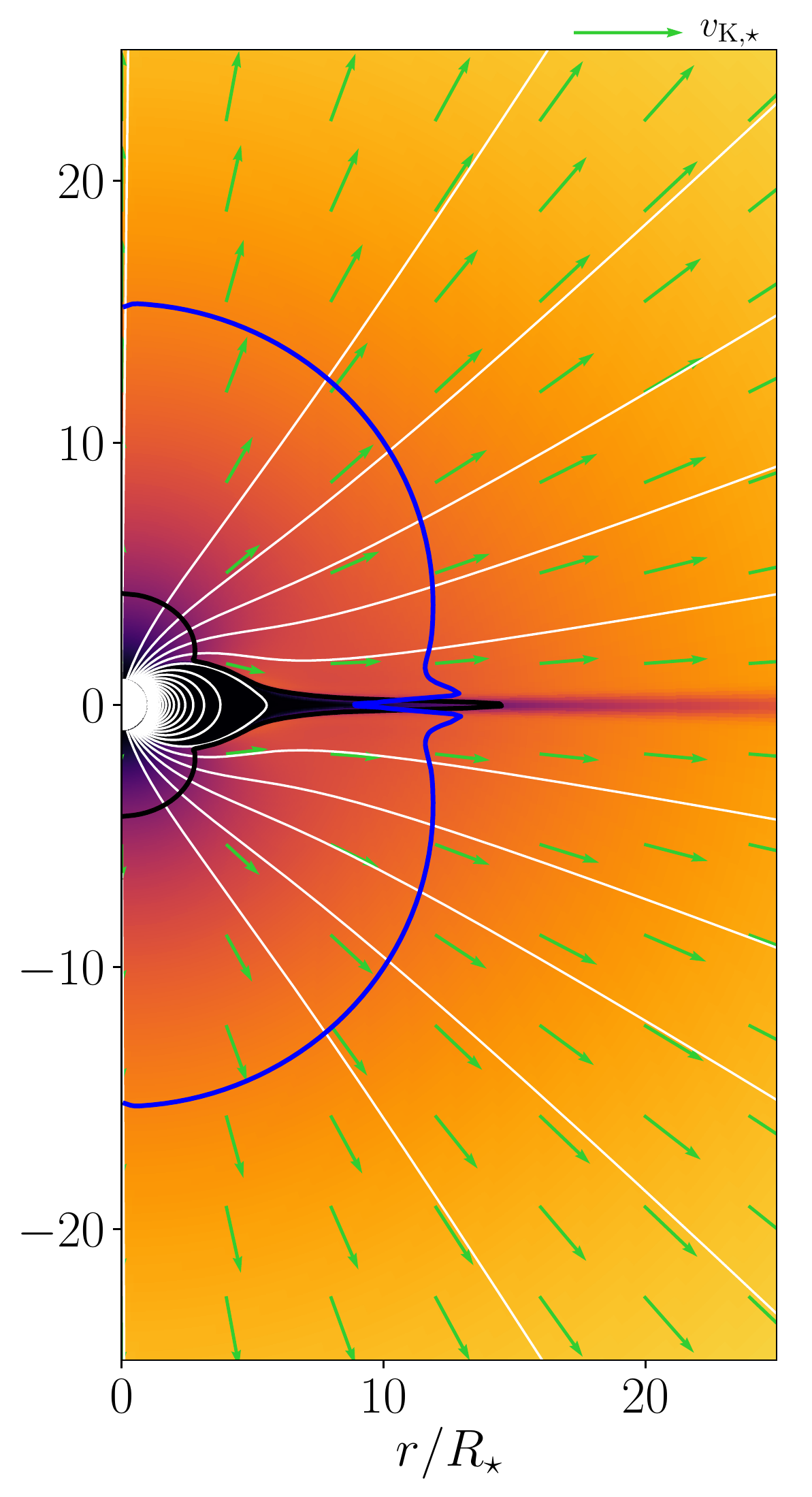}
\includegraphics[width=0.34155\textwidth]{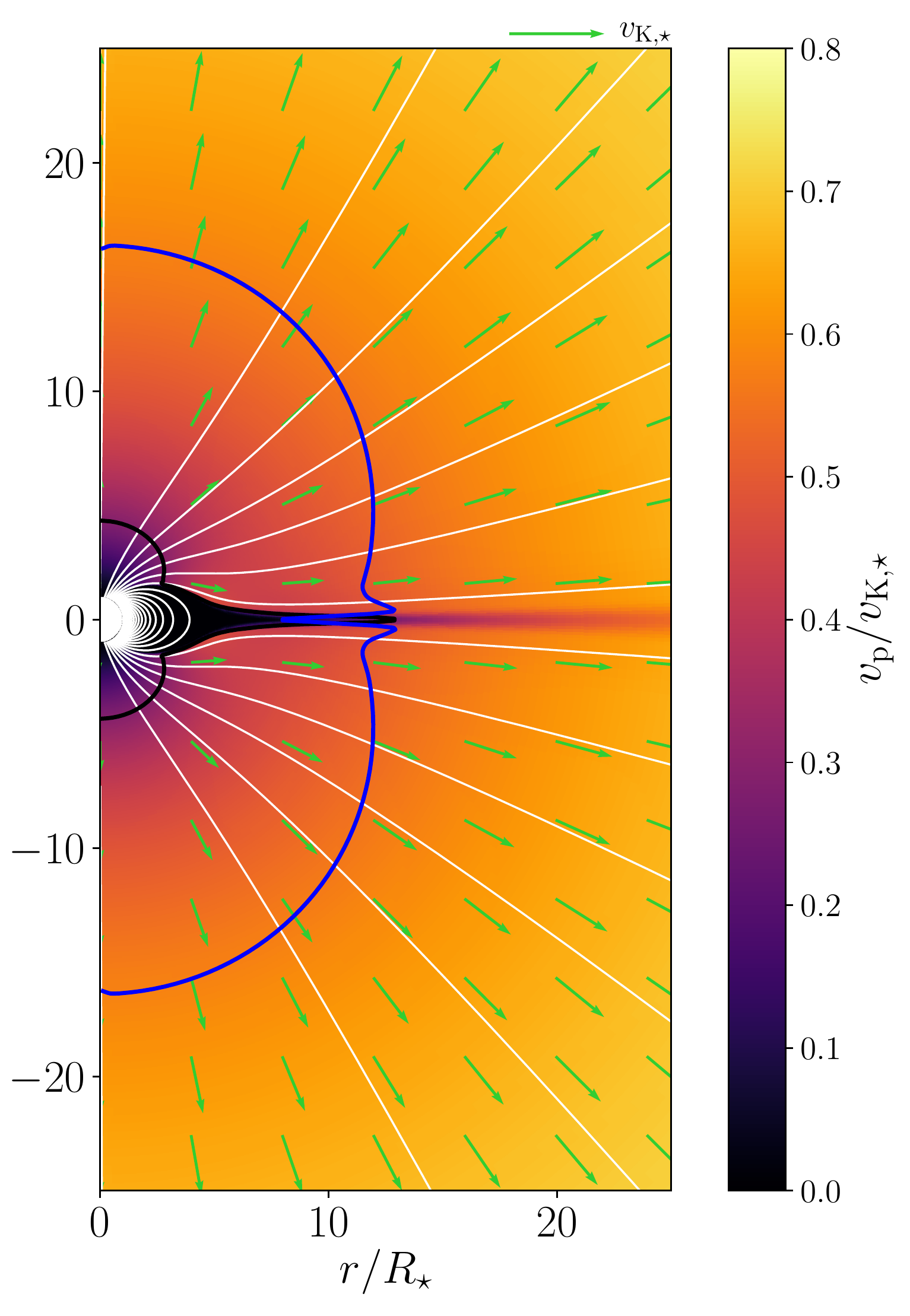}
\raisebox{-0.0085\height}{\includegraphics[width=0.079\textwidth]{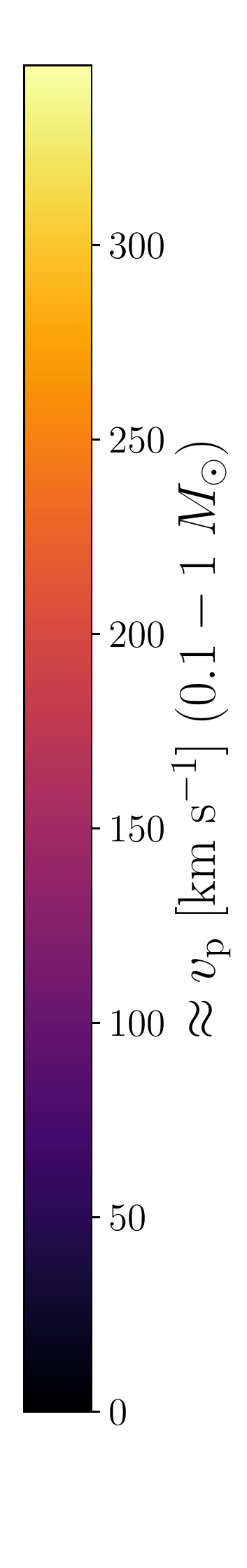}}

\caption{Poloidal velocity color maps, showing (left) model 46 ($f_\text{eq}=0.05$, $\alpha=0$), (center) model 14 ($f_\text{eq}=0.001$, $\alpha=0$), and (right) model 34 ($f_\text{eq}=0.05$, $\alpha=0.75$). All simulations above have a surface polar field strength $B_\star/B_0=2.83$. Magnetic field lines (white), velocity vectors normalized by $v_{\text{K},\star}$ (green), the sonic surface (black), and the Alfv\'en surface (blue) are included. {We include a secondary colorbar that gives poloidal velocity in physical units (km s$^{-1}$) for the $0.1 - 1$ $M_\odot$ cases in Table~\ref{tab:norm_sims}. As $v_{\text{K},\star}$ for these cases only varies by a maximum of $\approx 7 \%$ from the mean value $\langle v_{\text{K},\star} \rangle = 432.75$ km s$^{-1}$, we adopt this to produce a single colorbar that approximately represents the entire mass range.} \label{fig:frames_Vpol_150}}
\end{center}
\end{figure*}

In this section, we demonstrate the qualitative properties of these stellar wind simulations, focusing on the effects differential rotation have on the dynamical behavior of the stellar wind. Figure~\ref{fig:frames_Vpol_150} shows the poloidal velocity distribution for the computational domains of three stellar wind simulations: model 46 ($f_\text{eq}=0.05$, $\alpha=0$); model 14 ($f_\text{eq}=0.001$, $\alpha=0$); and model 34 ($f_\text{eq}=0.05$, $\alpha=0.75$). These are taken at simulation times corresponding to the middle of their respective time-median domains. The figures show the magnetic field lines (white), the ejection of stellar wind (green) along open field lines, the sonic Mach surface (black), and the Alfv\'en Mach surface (blue). For the rapid solid-body rotator (model 46), centrifugal effects become increasingly noticeable in the stellar wind, enhancing wind speeds at mid-to-high colatitudes and causing the wind to self-collimate and bend towards the poles; this results in an elongated Alfv\'en surface \citep[see, e.g.,][]{1985A&A...152..121S,1993MNRAS.262..936W,Matt_2004}. For the slow solid-body rotator (model 14), centrifugal effects are negligible, therefore the dynamics of the wind are predominantly determined by the stellar field strength and the mass-loss rate in the wind; this results in a more spherical Alfv\'en surface. For the differential rotator (model 34), even though the equator rotates as quickly as the rapid solid body case, the rotation of the wind itself (originating from low colatitudes) is much slower, reducing centrifugal effects and resulting in dynamical behavior similar to that of the slow solid-body case.

% \afterpage{
% \afterpage{
\begin{figure*}
\begin{center}
\includegraphics[width=0.45\textwidth]{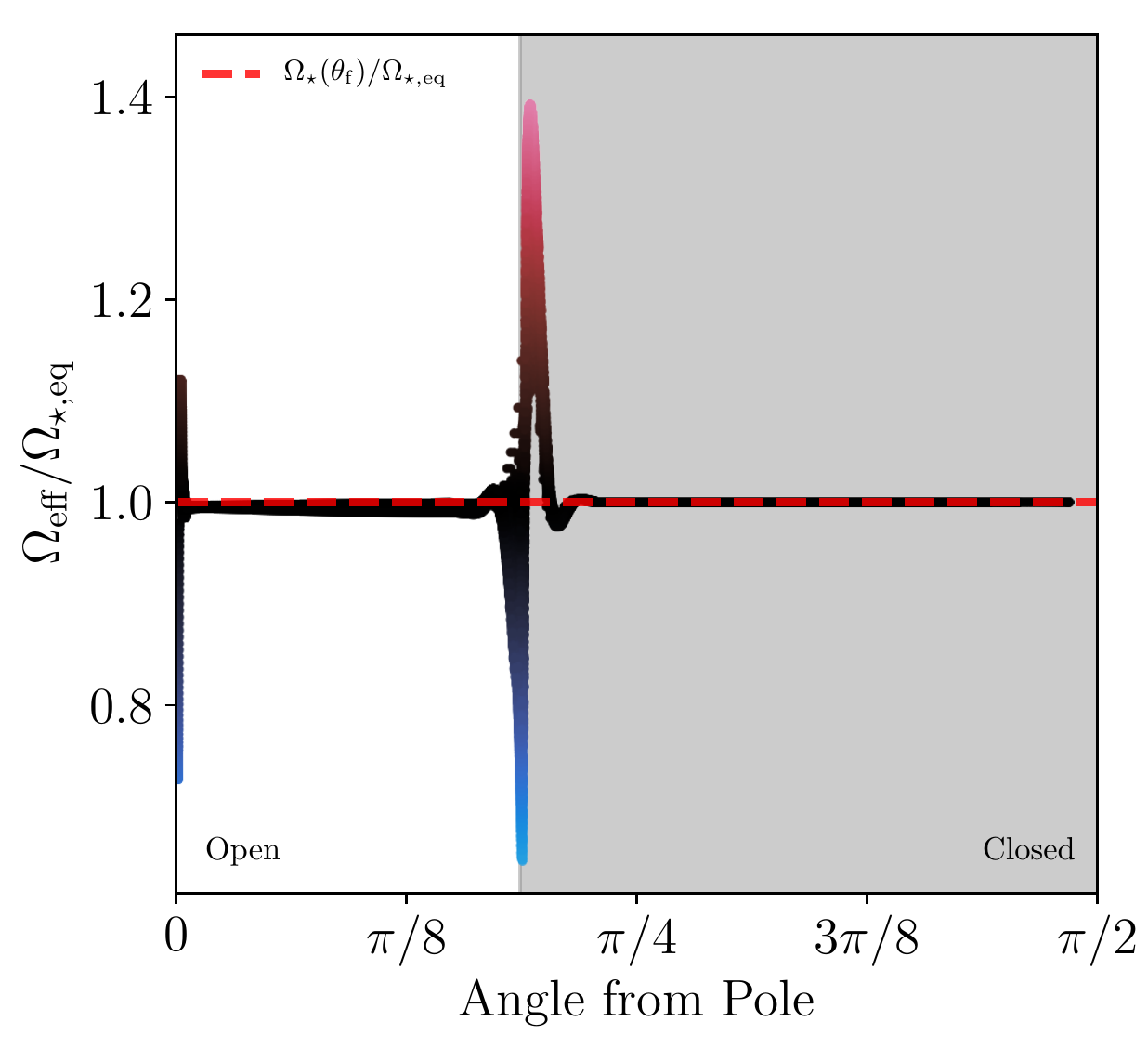}
\includegraphics[width=0.53\textwidth]{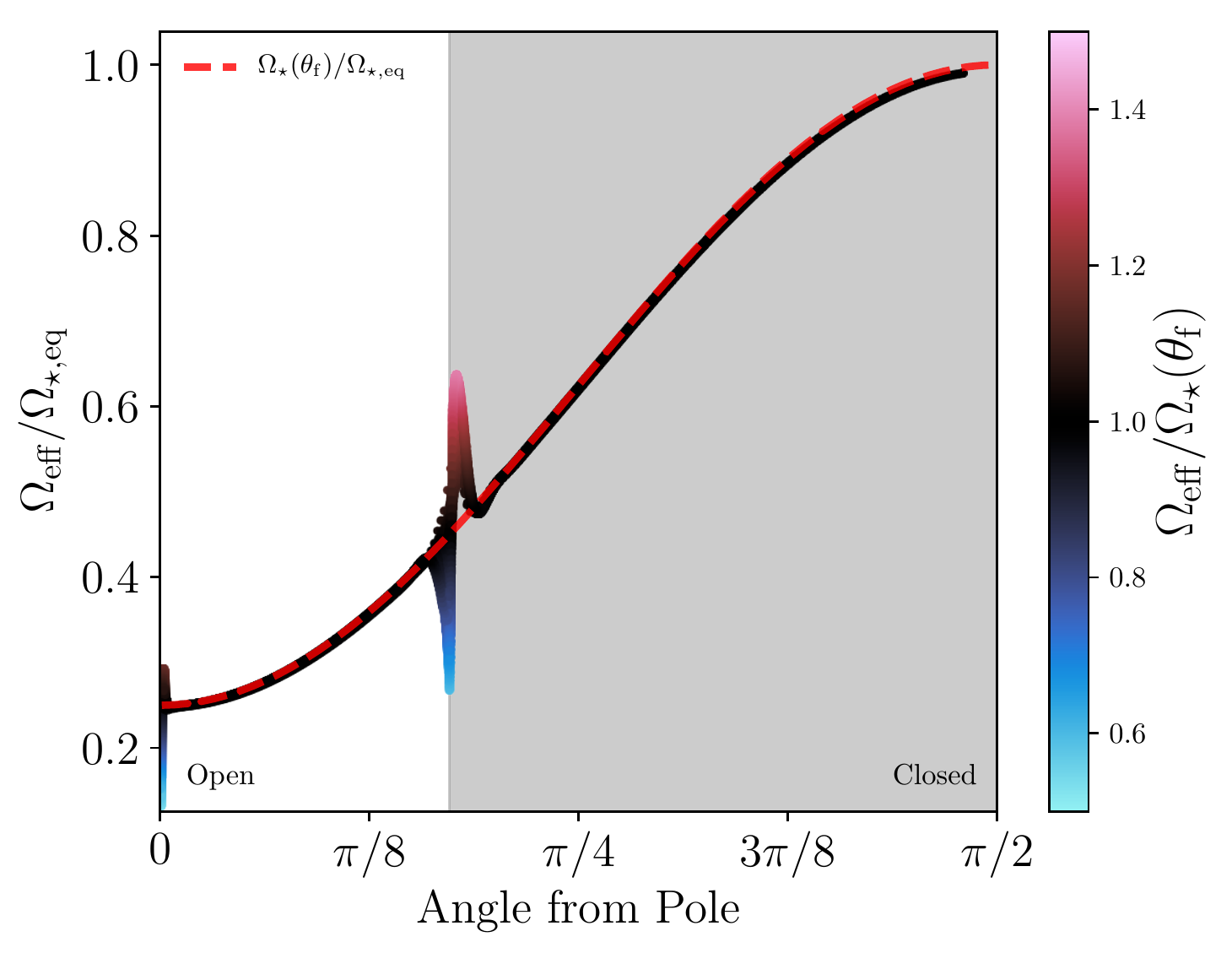}

\caption{$\Omega_\text{eff}/\Omega_{\star,\text{eq}}$ as a function of the angle on the stellar surface between the footprint colatitude and its closest pole, showing (left) model 46 ($f_\text{eq}=0.05$, $\alpha=0$) and (right) model 34 ($f_\text{eq}=0.05$, $\alpha=0.75$). Both simulations have a field strength $B_\star/B_0=2.83$. Red dotted line shows the stellar differential rotation profile $\Omega_{\star}(\theta_\text{f})$ adopted at the inner boundary of the computational domain (Equation~(\ref{eq:DR_profile})), normalized by $\Omega_{\star,\text{eq}}$. Color map illustrates divergence of $\Omega_\text{eff}$ from the stellar differential rotation profile $\Omega_{\star}(\theta_\text{f})$. White (gray) area illustrates the region on the stellar surface in which the magnetic field is open (closed).  \label{fig:Omegaeff_div_star_eq_v_theta}}
\end{center}
\end{figure*}

% }
% }
In addition, we probe differences in the rotation of the wind between solid body and differential rotation models, by calculating the effective rotation rate across the computational domain:

\begin{align}\label{eq:eff_omega}
\begin{split}
\Omega_\text{eff} = \frac{1}{R \sin{\theta}} \left(v_\phi - \frac{B_\phi \kappa}{\rho} \right).
\end{split}
\end{align}
In order to calculate accurate stellar torque values, this quantity is required to be conserved within some tolerance along magnetic field lines \citep[as shown in][]{2009A&A...508.1117Z}. In practice, we check its convergence by calculating $\Omega_\text{eff}$ in each grid cell and tracing the field line at this point back to its footpoint on the stellar surface. We then compare this quantity to the stellar surface rotation rate at that given footpoint's colatitude, $\theta_\text{f}$, i.e., $\Omega_{\star}(\theta_\text{f})$; full conservation of this quantity is satisfied when $\Omega_\text{eff}(R,\theta)=\Omega_{\star}(\theta_\text{f})$. In Figure~\ref{fig:Omegaeff_div_star_eq_v_theta}, we plot $\Omega_\text{eff}$ (normalized by the equatorial spin rate, $\Omega_{\star,\text{eq}}$) as a function of the angle on the stellar surface between the footprint colatitude and its closest pole, for the same rapid solid-body (model 46) and differential rotation (model 34) cases from Figure~\ref{fig:frames_Vpol_150}. We plot the differential rotation profile (expressed in Equation~(\ref{eq:DR_profile})) as a function of $\theta_\text{f}$ (normalized by $\Omega_{\star,\text{eq}}$) for comparison, where the color map illustrates the divergence of $\Omega_\text{eff}$ from this profile.\footnote{Divergence between $\Omega_\text{eff}$ and $\Omega_{\star}(\theta_\text{f})$ typically occurs at the transition between open and closed field lines, due to increasing numerical diffusion.} In both cases, the white area represents the area on the stellar surface in which the stellar wind is being ejected, i.e., where the magnetic field lines are open.  It is evident that the stellar wind region decreases with increasing differential rotation. For solid-body rotation, $\Omega_\text{eff}$ appears to be well-conserved in the region of the stellar wind (where $\Omega_\text{eff}/\Omega_{\star}(\theta_\text{f})\approx 1$), and the rotation of the wind is roughly constant. For differential rotation, $\Omega_\text{eff}$ is also well-conserved in the wind, but the rotation of the wind decreases with decreasing colatitude (towards the poles).

% \afterpage{
% \afterpage{
% }
% }
\subsection{Global Flow Quantities and Efficiencies}\label{sec:glob_flow}

The main aim of this paper is to investigate how the stellar wind interacts with a star that is undergoing differential rotation. By integrating over a spherical shell of surface $S$ perpendicular to the poloidal flow at a given radius $R$, we express the radial profiles  of the stellar wind mass flow rate, torque, and unsigned magnetic flux as

\begin{equation}\label{eq:mass_rate_int}
\dot{M} = \int_{S} \rho \boldsymbol{v_\text{p}} \cdot \text{d}\boldsymbol{S},
\end{equation}

\begin{align}\label{eq:torque_int}
\begin{split}
\dot{J} = \int_{S} \Lambda \rho \boldsymbol{v_\text{p}} \cdot \text{d}\boldsymbol{S},
\end{split}
\end{align}
and

\begin{equation}\label{eq:unsigned_flux}
\Phi = \int_{S} \lvert\boldsymbol{B} \cdot \text{d}\boldsymbol{S}\rvert,
\end{equation}
respectively, where

\begin{align}\label{eq:AM}
\begin{split}
\Lambda = R \sin{\theta} \left(v_\phi - \frac{B_\phi}{\kappa} \right)
\end{split}
\end{align}
is a quantity related to the angular momentum flux $\boldsymbol{F_\text{AM}}=\Lambda \rho \boldsymbol{v}$. In practice, we determine global values of $\dot{M}$ and $\dot{J}$ for all simulations by taking the median values over all spherical shells at $R > 7.68 R_\star$ (corresponding to the outer half of the logarithmic radial grid), in order to avoid numerical effects near the inner boundary. The unsigned magnetic flux for a system with dipolar topology initially falls as $1/R$, but reaches a constant value that represents the open flux in the stellar wind, $\Phi_\text{open}$, outside the Alfv\'en surface, i.e., when the thermal and hydrodynamic wind pressure exceeds the magnetic pressure. In practice, we determine global values of $\Phi_\text{open}$ for all simulations by taking the median values over all spherical shells at $R > 40 R_\star$. Values of $\dot{M}$, $\dot{J}$, and $\Phi_\text{open}$ for all simulations can be found in Table~\ref{tab:Pluto_sims}.

\begin{figure}
\begin{center}
\includegraphics[width=0.4625\textwidth]{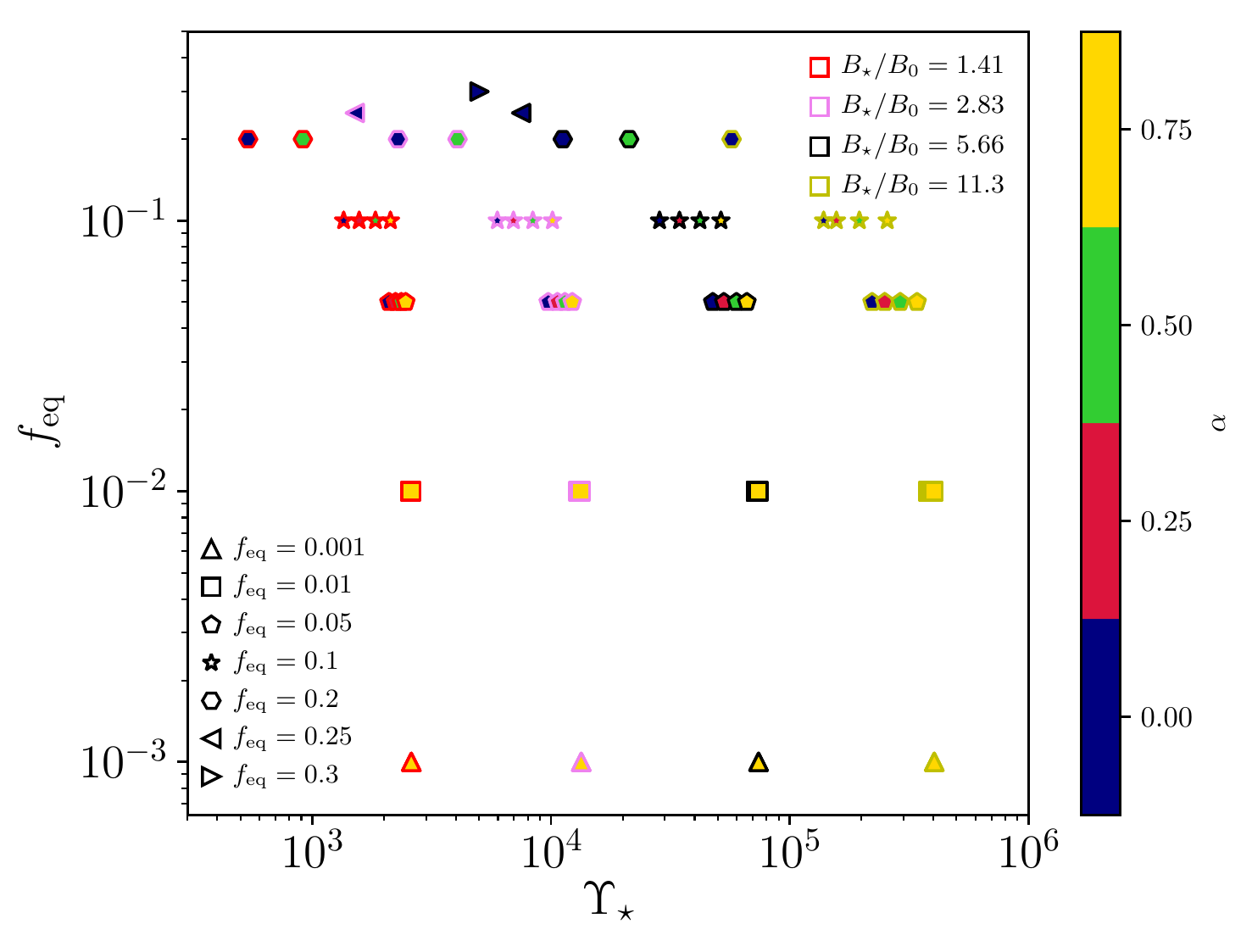}
\caption{$f_\text{eq}$ as a function of $\Upsilon_\star$, for all simulations. To differentiate between different input parameters, we use differing symbol styles for $f_\text{eq}$, marker colors for $\alpha$, and marker edge colors for $B_\star$. \label{fig:f_0_v_Upsilon}}
\end{center}
\end{figure}

We define a ``wind magnetization" parameter that encapsulates the magnetic and mass-loss stellar wind properties, based on properties at the stellar surface:

\begin{equation}\label{eq:Y_star}
\Upsilon_\star = \frac{\Phi_\star^2}{4 \pi R_\star^2\dot{M} v_\text{esc}},
\end{equation}
where $\Phi_\star = \alpha \pi R_\star^2 B_\star$ is the total stellar flux ($\alpha=2$ for our dipolar surface configuration). Values of $\Upsilon_\star$ for all simulations can be found in Table~\ref{tab:Pluto_sims}. In Figure~\ref{fig:f_0_v_Upsilon}, we illustrate the extent of the parameter space explored by our simulations, in terms of $f_\text{eq}$ and $\Upsilon_\star$. Symbol shapes, colors, and borders demonstrate how each parameter in our study varies. Increasing $B_\star$ noticeably increases $\Upsilon_\star$ (at fixed $f_\text{eq}$ and $\alpha$). For slower rotation rates ($f_\text{eq} \leq 0.01$), changes in $f_\text{eq}$ and/or $\alpha$ have a negligible effect on $\Upsilon_\star$ (at fixed $B_\star$); therefore, simulations with changing $\alpha$ (for fixed $B_\star$ and $f_\text{eq}$) are seen to overlap in this figure. For more rapid rotation ($f_\text{eq}\geq 0.05$), increasing $f_\text{eq}$ and/or decreasing $\alpha$ decreases $\Upsilon_\star$ (at fixed $B_\star$).

% \afterpage{
% }

{Equation~(\ref{eq:Y_star}) demonstrates that these simulations are degenerate in $B_\star^2/\dot{M}$. This means that for a real star, the range of $\Upsilon_\star$ shown in Figure~\ref{fig:f_0_v_Upsilon} can be taken to represent a range of $B_\star$ for some fixed value of $\dot{M}$, a range of $\dot{M}$ for some fixed $B_\star$, or for any combination of the two. Table~\ref{tab:norm_sims} shows the range of $B_\star$ and $\dot{M}$ in physical units, for a few representative MS star masses and choices of coronal-density normalization. Similarly, the range of $f_\text{eq}$ shown in Figure~\ref{fig:f_0_v_Upsilon} represent a range of equatorial rotation periods, where example physical values are given in Table~\ref{tab:norm_sims}.}

\section{Torque Formulation for Simulations of Stellar Winds}\label{sec:torque_formulation}

Through dimensional analysis, a general parameterization for the stellar wind torque can be written as

\begin{equation}\label{eq:torque_wind}
\dot{J} = \dot{M} \langle \Omega_{\text{wind}} \rangle \langle r_\text{A} \rangle^2,
\end{equation}
where $\langle \Omega_{\text{wind}} \rangle$ represents an ``effective" stellar rotation rate in the stellar wind region, and $\langle r_\text{A} \rangle$ is a characteristic length scale that represents the ``effective magnetic lever arm" and quantifies the efficiency of the stellar wind torque \citep[see, e.g.,][]{1967ApJ...148..217W}. The Alfv\'en radius (normalized by $R_\star$) is determined by rearranging Equation~(\ref{eq:torque_wind}):
\begin{equation}\label{eq:RA_torque_wind}
\frac{\langle r_\text{A} \rangle}{R_\star} = \left(\frac{\dot{J} }{ \dot{M} \langle \Omega_\text{wind} \rangle R_\star^2}\right)^{1/2}
\end{equation}
\citep[following][]{1993MNRAS.262..936W,2008ApJ...678.1109M}. 
However, $\langle \Omega_{\text{wind}} \rangle$ is not known a priori, so we instead normalize the angular momentum loss in the stellar wind using the equatorial rotation rate $\Omega_{\star,\text{eq}}$, giving

\begin{equation}\label{eq:RAeq_torque_wind}
\frac{\langle r_\text{A,eq} \rangle}{R_\star} = \left(\frac{\dot{J} }{ \dot{M} \Omega_{\star,\text{eq}} R_\star^2}\right)^{1/2} \equiv \omega^{1/2} \frac{\langle r_\text{A} \rangle}{R_\star},
\end{equation}
where we define

\begin{equation}\label{eq:omega}
\omega = \frac{\langle \Omega_{\text{wind}} \rangle}{\Omega_{\star,\text{eq}}} \equiv \frac{\langle f_{\text{wind}} \rangle}{f_\text{eq}},
\end{equation}
where $\langle f_{\text{wind}} \rangle$ represents an ``effective" stellar break-up fraction in the stellar wind. Values of $\langle r_\text{A,eq} \rangle/R_\star$ for all simulations can be found in Table~\ref{tab:Pluto_sims}. Therefore, we rewrite Equation~(\ref{eq:torque_wind}) in terms of simulation parameters:
\begin{align}\label{eq:torque_wind_eq}
\begin{split}
\dot{J} = \dot{M} \Omega_{\star,\text{eq}} \langle r_\text{A,eq} \rangle^2,
\end{split}
\end{align}

We adopt the stellar wind torque formulation developed by \citet{2012ApJ...754L..26M}:

\begin{equation}\label{eq:Ra_Y_star}
\frac{\langle r_\text{A}\rangle}{R_\star} = K_{\star,1} \left(\frac{\Upsilon_\star}{\beta_\star}\right)^{m_\star},
\end{equation}
where $\Upsilon_\star$ is the ``magnetization parameter'' of the stellar wind based on properties at the stellar surface (expressed in Equation~(\ref{eq:Y_star})), 

\begin{equation}\label{eq:beta_star}
\beta_\star = \left[1 + \left(\frac{\omega f_\text{eq}}{K_{\star,2}}\right)^2\right]^{1/2}
\end{equation}
is a ``magnetocentrifugal correction" term for the wind efficiency (expressed in the form adopted by \citealt{2015ApJ...798..116R}), and $K_{\star,1}$, $K_{\star,2}$, and $m_\star$ are best-fit dimensionless parameters. The Alfv\'en radius is inversely proportional to the stellar wind mass-loss rate. Using Equation~(\ref{eq:RAeq_torque_wind}), we rewrite Equation~(\ref{eq:Ra_Y_star}) in terms of $\langle r_\text{A,eq} \rangle/R_\star$:

\begin{equation}\label{eq:Ra_Y_star_eq}
\frac{\langle r_\text{A,eq}\rangle}{R_\star} = K_{\star,1} \left(\frac{\Upsilon_\star}{\beta_\star}\right)^{m_\star} \omega^{1/2}.
\end{equation}

\begin{figure}
\begin{center}
\includegraphics[width=0.4625\textwidth]{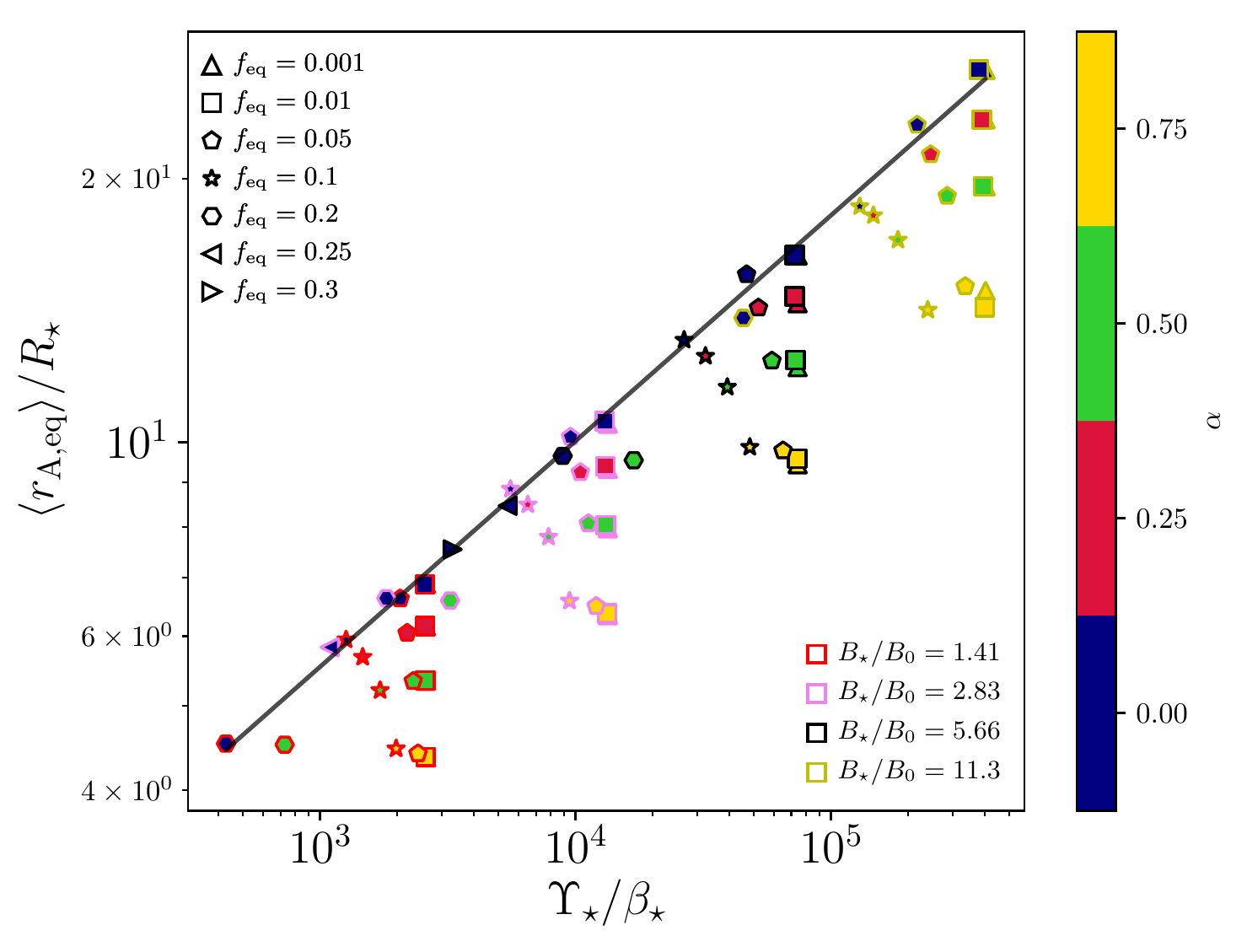}
\caption{$\langle r_\text{A,eq} \rangle/R_\star$ as a function of $\Upsilon_\star/\beta_\star$ (fixing $\omega = 1$), for all simulations. Black line shows the solid body  fit $\langle r_\text{A,eq} \rangle/R_\star = 0.932 (\Upsilon_\star / \beta_\star)^{0.258}$ (with $K_{\star,2}=0.263$). \label{fig:rA_v_Upsilon_div_beta_star}}
\end{center}
\end{figure}

In Figure~\ref{fig:rA_v_Upsilon_div_beta_star}, we plot $\langle r_\text{A,eq}\rangle/R_\star$ as a function of $\Upsilon_\star/\beta_\star$, (fixing $\omega = 1$ for the $\beta_\star$ term). 
By solely fitting the solid-body rotating cases, we find the best-fit dimensionless parameters to be $K_{\star,1}=0.932$, $K_{\star,2}=0.263$, and $m_\star=0.258$. For convenience, all of the best-fit dimensionless parameters presented in Section~\ref{sec:torque_formulation} can be found in Table~\ref{tab:dimensionless_params}. 
The cases with solid-body rotation are well fit by the power-law represented by Equation~(\ref{eq:Ra_Y_star}), illustrated by the small amount of scatter of those cases around the best-fit line.
However, the cases with differential rotation exhibit a systematic deviation of the points toward smaller $\langle r_\text{A,eq}\rangle/R_\star$ values for larger amounts of differential rotation (larger $\alpha$).
It is clear that any stellar wind torque formulation that assumes solid-body rotation (i.e., one that uses the equatorial rotation rate and assumes $\omega=1$) will increasingly overestimate the efficiency of the stellar wind torque for increasing solar-like differential rotation. 
We can define an empirical value for the normalized ``effective" rotation rate in the wind, $\omega_\text{emp}$, as being the deviation of the simulated values from the simple power-law relationship of the solid-body-rotator case.  That is, the ``effective" rotation rate in the wind can be measured using Equation~(\ref{eq:Ra_Y_star_eq}), and the fit parameters ($K_{\star,1}$, $K_{\star,2}$, and $m_\star$): 

\begin{equation}\label{eq:omega_empirical}
\omega_\text{emp} = \left(\frac{\langle r_\text{A,eq}\rangle/R_\star}{K_{\star,1} (\Upsilon_\star/\beta_\star)^{m_\star}}\right)^{2}.
\end{equation}
Equation~(\ref{eq:omega_empirical}) is a transcendental equation (due to the $\omega$ factor present in $\beta_\star$), therefore we solve this iteratively. Values of $\omega_\text{emp}$ for all simulations can be found in Table~\ref{tab:Pluto_sims}.

\begin{deluxetable}{cccc}
\tablecaption{Best-fit Dimensionless Parameters from Scaling Laws in Section~\ref{sec:torque_formulation}.\label{tab:dimensionless_params}}

\tablehead{Parameter & Value & Equations}

\startdata
$K_{\star,1}$ & 0.932  & (\ref{eq:Ra_Y_star}) \\
$K_{\star,2}$ & 0.263  & \\
$m_\star$ & 0.258  & \\
$K_\theta$ & 1.06 & (\ref{eq:omega_full}) \\
$K_{\Phi,1}$ & 1.16 & \\
$K_{\Phi,2}$ & 0.0111 & \\
$m_\Phi$ & -0.165  &
\enddata
\end{deluxetable}

In order to predict the stellar wind torque as a function of the stellar surface properties, we must adopt a formulation for $\omega$. 
The ``effective" rotation rate in the wind, as defined by $\omega$, is a torque-averaged rotation rate over all open field lines carrying wind material.
Since our adopted differential rotation profile is a simple function of colatitude on the stellar surface, we can define the colatitude on the star that happens to rotate at the same rate as the ``effective" rotation of the wind, $\theta_\text{eff}$.
Equations~(\ref{eq:DR_profile}) and (\ref{eq:omega}) thus define the ``effective" rotation rate in terms of this colatitude:
\begin{align}\label{eq:omega_2a}
\begin{split}
\omega_\text{th} {}& = 1 - \alpha \cos^2{\theta_\text{eff}}.
\end{split}
\end{align}
Since the rotation rate increases from pole to equator, and since the integrated torque (and thus ``effective" rotation in the wind) is dominated by flow from the highest colatitudes, we expect that $\theta_\text{eff}$ will have a similar value and similar scaling to the maximum colatitude on the stellar surface that emits a wind (as opposed to the higher colatitudes where there is a dead zone), which we call $\theta_\text{open}$.
In practice, we explored a few different possible relationships between $\theta_\text{eff}$ and $\theta_\text{open}$.  
Given the functional form of Equation~(\ref{eq:omega_2a}), we found that a relatively simple form and acceptable fit to the data is obtained by assuming that $\cos^2{\theta_\text{eff}}$ is proportional to $\cos^2{\theta_\text{open}}$, leading to
\begin{align}\label{eq:omega_2}
\begin{split}
\omega_\text{th} {}& = 1 - \alpha K_\theta \cos^2{\theta_\text{open}},
\end{split}
\end{align}
where $K_\theta$ is a best-fit dimensionless parameter.
By evaluating Equation~(\ref{eq:unsigned_flux}) at $R_\star$ across the average area enclosing the stellar wind, i.e., for $0 \leq \theta \leq \theta_\text{open}$ and $(\pi - \theta_\text{open}) \leq \theta \leq \pi$, using the dipolar definition for $B_R$ (Equation~(\ref{eq:Br})), one can write

\begin{equation}\label{eq:phi_wind}
\cos^2{\theta_\text{open}} = 1 - \frac{\Phi_\text{open}}{\Phi_\star},
\end{equation}
giving

\begin{equation}\label{eq:omega_3}
\omega_\text{th} =1 - \alpha K_\theta \left(1 - \frac{\Phi_\text{open}}{\Phi_\star}\right)
\end{equation}
for our dipolar configuration. Therefore, when $\alpha \neq 0$, the ``effective" rotation rate of the wind (normalized by the equatorial stellar rotation rate) is a function of $\alpha$ and the fraction of open flux in the stellar wind.

In previous stellar wind simulations, it has been shown that the fraction of open flux increases with the wind mass-loss rate (at fixed $f_\text{eq}/B_\star$); more specifically, $\Phi_\text{open}/\Phi_\star$ scales inversely with $\Upsilon_\star$ \citep[see, e.g.,][]{2015ApJ...798..116R,2017ApJ...845...46F,2018ApJ...854...78F,2017ApJ...849...83P}. Therefore, we parameterize $\Phi_\text{open}/\Phi_\star$ using the following functional fit:

\begin{equation}\label{eq:phi_w_param}
\frac{\Phi_\text{open}}{\Phi_\star} = K_{\Phi,1} \left(\frac{\Upsilon_\star}{\beta_\Phi}\right)^{m_\Phi},
\end{equation}
where

\begin{equation}\label{eq:beta_phi}
\beta_\Phi = \left\{1 + \left[\frac{(1-\alpha) f_\text{eq}}{K_{\Phi,2}}\right]^2\right\}^{1/2}
\end{equation}
is the ``magnetocentrifugal correction" term for the fraction of open flux in the stellar wind, and $K_{\Phi,1}$, $K_{\Phi,2}$, and $m_\Phi$ are best-fit dimensionless parameters. By solely fitting the solid-body rotation cases, we find the best-fit dimensionless parameters to be $K_{\Phi,1}=1.16$, $K_{\Phi,2}=0.0111$, and $m_\Phi=-0.165$.\footnote{Note, $K_{\Phi,2}\neq K_{\star,2}$. The $K_{\star,2}$ term captures the combined effects of the wind acceleration and the fact that the open flux is changing, whereas $K_{\Phi,2}$ captures only the latter.} In Equation~(\ref{eq:phi_w_param}), we assume $\omega \approx (1-\alpha)$ in the $\beta_\Phi$ term for simplicity, to avoid producing a transcendental equation for the full parameterization of $\omega$; this approximation gives simulation points that lie within $\lesssim 10 \%$ of the formulation if $\omega$ was used. Therefore, one can rewrite Equation~(\ref{eq:omega_3}) as

\begin{equation}\label{eq:omega_full}
\omega_\text{th} = 1 - \alpha K_\theta \left[1 - K_{\Phi,1} \left(\frac{\Upsilon_\star}{\beta_\Phi}\right)^{m_{\Phi}}\right].
\end{equation}

We determine the best-fit dimensionless parameter $K_\theta = 1.06$, by fitting $\omega_\text{emp}$ values with the expression for $\omega_\text{th}$ (Equation~(\ref{eq:omega_full})). 
Conceptually, given the differential profile used in this study, $\theta_\text{open}$ corresponds to the colatitude of the field line with the maximum stellar rotation rate in the stellar wind.
Our value of $K_\theta$ demonstrates that $\langle \Omega_{\text{wind}} \rangle$ acts along a field line connected at a colatitude that is near to, but lower than, $\theta_\text{open}$.
In other words, the location on the stellar surface that rotates at the same rate as the ``effective" rotation rate of the wind is located within the wind-emitting region and near the last open field line, as expected.

\begin{figure}
\begin{center}
\includegraphics[width=0.4625\textwidth]{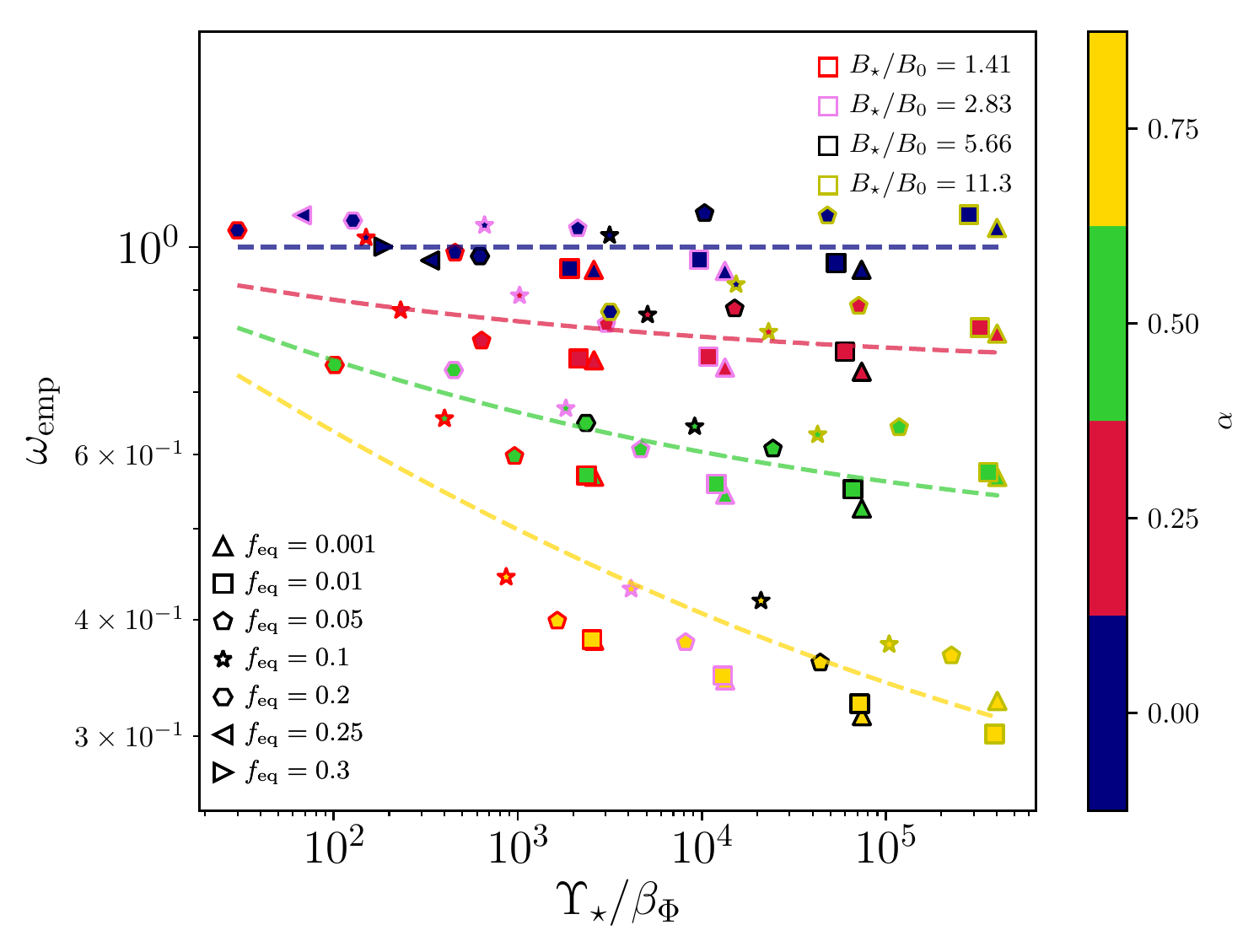}
\caption{$\omega_\text{emp}$ as a function of $\Upsilon_\star/\beta_\Phi$, for all simulations. Dotted lines represent the theoretical solution $\omega_\text{th}$, for a given $\alpha$. \label{fig:omega_v_Upsilon_div_beta_Phi}}
\end{center}
\end{figure}

In Figure~\ref{fig:omega_v_Upsilon_div_beta_Phi}, we plot $\omega_\text{emp}$ as a function of $\Upsilon_\star/\beta_\Phi$. When the differential rotation increases, i.e., as $\alpha$ increases, the ``effective" rotation rate in the wind is smaller. For a given differential rotation profile, the trend with $\Upsilon_\star/\beta_\Phi$ illustrates that the ``effective" rotation rate in the wind is sensitive to the fraction of open flux in the wind. Specifically, if there is more fractional open flux (which occurs for smaller $\Upsilon_\star/\beta_\Phi$), the ``effective" rotation of the wind is faster, because it includes more flow from higher colatitudes. We superimpose lines representing the fully parameterized $\omega_\text{th}$ for each $\alpha$ explored in our study (Equation~(\ref{eq:omega_full})). Scatter present for $\alpha=0$ cases are solely due to deviations in the relationship between $\langle r_\text{A,eq} \rangle/R_\star$ and $\Upsilon_\star/\beta_\star$, whereas for $\alpha>0$ cases, scatter is likely a combination of this and from our parameterization of $\omega$.

\begin{figure}
\begin{center}
\includegraphics[width=0.4625\textwidth]{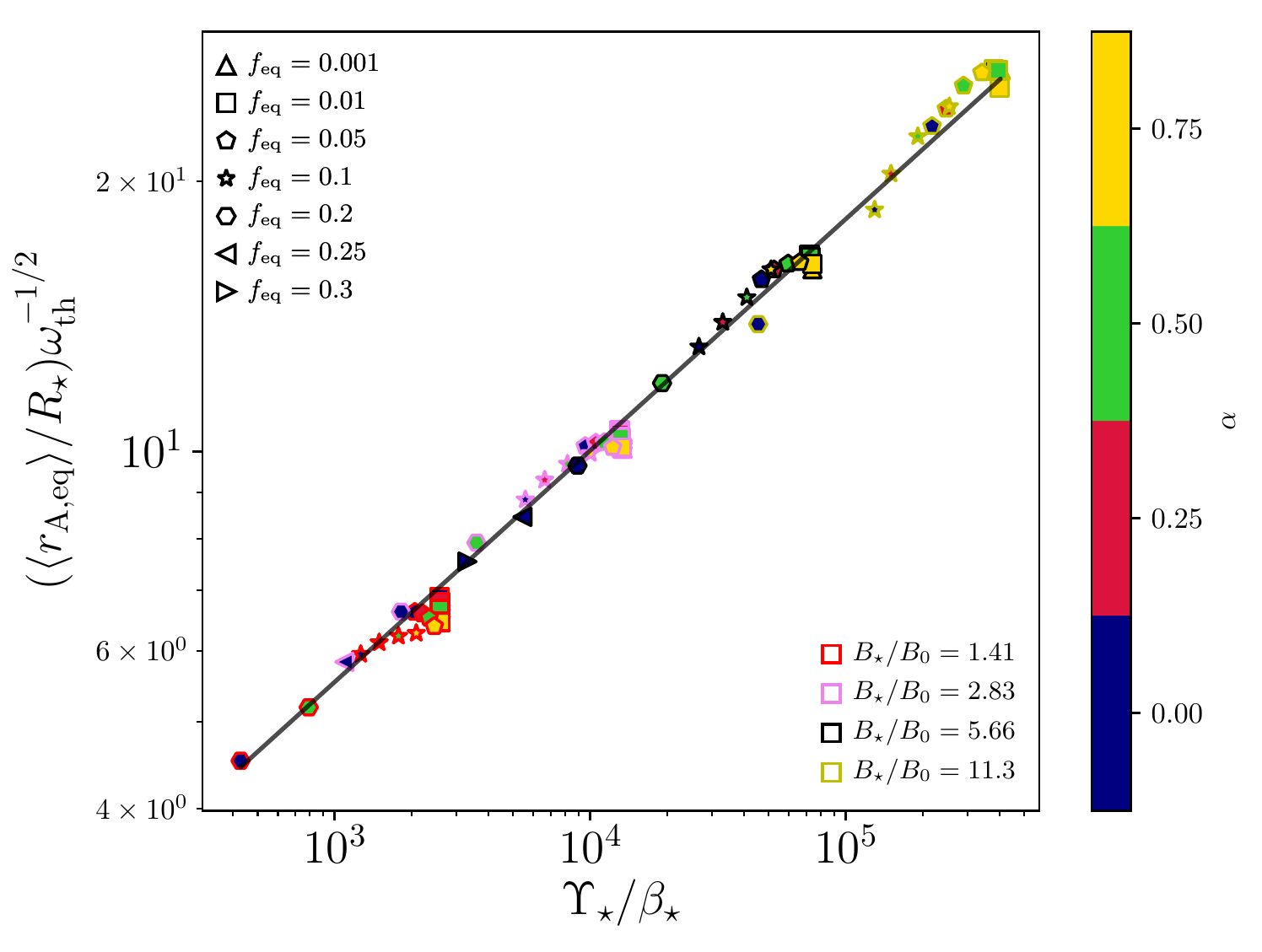}
\caption{$(\langle r_\text{A,eq} \rangle/R_\star) \omega_\text{th}^{-1/2}$ as a function of $\Upsilon_\star / \beta_\star$, for all simulations. Black line shows the fit $(\langle r_\text{A,eq} \rangle/R_\star) \omega_\text{th}^{-1/2} = 0.932 (\Upsilon_\star / \beta_\star)^{0.258}$ (with $K_{\star,2}=0.263$). \label{fig:rA_div_omega_v_Upsilon_div_beta_star}}
\end{center}
\end{figure}

In Figure~\ref{fig:rA_div_omega_v_Upsilon_div_beta_star}, we plot $(\langle r_\text{A,eq}\rangle/R_\star) \omega_\text{th}^{-1/2}$ as a function of $\Upsilon_\star/\beta_\star$ (using $\omega_\text{th}$ in the expression of $\beta_\star$). We find all simulation points to lie within $\lesssim 10 \%$ of the fitting function. Therefore, our adoption (and parameterization) of $\omega$ appears to strongly account for the stellar differential rotation effects on the efficiency of the stellar wind torque. Finally, by combining Equations~(\ref{eq:torque_wind_eq}), (\ref{eq:Ra_Y_star_eq}), and (\ref{eq:omega_full}), we express a fully parameterized estimate for the stellar wind torque, for a wide range of differential rotation rates and stellar surface field strengths:

\begin{align}\label{eq:torque_wind_full_param}
\begin{split}
\dot{J} = {}& K_{\star,1}^2 \dot{M} \Omega_{\star,\text{eq}} R_\star^2  \left(\frac{\Upsilon_\star}{\beta_\star}\right)^{2m_\star} \\
& \times \left\{1 - \alpha K_\theta \left[1 - K_{\Phi,1} \left(\frac{\Upsilon_\star}{\beta_\Phi}\right)^{m_{\Phi}}\right] \right\}.
\end{split}
\end{align}

\section{Discussion and Conclusions}\label{sec:disc_conc}

In this work, we perform 74 2.5D axisymmetric MHD stellar wind simulations to investigate how the stellar wind torque experienced by the star is impacted by differential rotation and the dipolar stellar surface magnetic field strength. We create a torque formulation that encapsulates these effects, allowing for the investigation of the rotational evolution of MS stars using 1D stellar evolution codes. We chose a simplified setup, including a purely dipolar magnetic field, thermal/polytropic wind driving, and a monotonic differential rotation profile.

\subsection{How Differential Rotation Affects Stellar Wind Torques}

We determine a stellar wind torque formulation through the parameterization of the effective Alfv\'en radius, based on the formulation employed by \citet{2012ApJ...754L..26M}. Our solid-body rotation simulations follow a $\langle r_\text{A} \rangle$-$\Upsilon_\star$-$f$ relationship (Equation~(\ref{eq:Ra_Y_star})) with a power-law index of $m_\star=0.258$, which is comparable to that found in previous stellar wind studies (with an initial dipolar configuration) within systematic uncertainties \citep[see, e.g.][]{2012ApJ...754L..26M,2015ApJ...798..116R,2017ApJ...845...46F,2017ApJ...849...83P,Pantolmos2020}. 
In addition, our magnetocentrifugal correction term for the wind efficiency, $\beta_\star$, has a large fitting constant of $K_{\star,2}=0.263$, suggesting that rotation has an impact on the wind acceleration only for our most-rapid cases. 
This value is slightly larger than that determined by \citet{2015ApJ...798..116R} ($\sim 0.2$), but much larger than that determined by \citet{2012ApJ...754L..26M} (0.07); as discussed in the former, these discrepancies might be due to differences in boundary conditions.

For cases undergoing surface differential rotation, the effective Alfv\'en radius now depends on the ``effective" stellar rotation rate in the wind, $\langle \Omega_\text{wind}\rangle$, which is not known a priori (Equation~(\ref{eq:RA_torque_wind})). We renormalize the Alfv\'en radius using the equatorial stellar rotation rate, $\Omega_{\star,\text{eq}}$, introducing the term $\omega=\langle \Omega_\text{wind}\rangle/\Omega_{\star,\text{eq}}$ into the formulation (Equation~(\ref{eq:RAeq_torque_wind})). We find that $\omega$ decreases with the magnitude of surface differential rotation (measured via $\alpha$). Furthermore, $\omega$ increases with the fraction of open flux in the wind (for a given $\alpha$); this corresponds to the stellar wind occupying a larger area of the domain, allowing for the inclusion of faster flow from higher colatitudes.
The fraction of open flux in the wind itself inversely scales with the wind magnetization, $\Upsilon_\star$; for rapidly-rotating cases, it is also sensitive to magnetocentrifugal effects on the wind acceleration and the open flux combined, $\beta_\Phi$. Our simulations follow a $\Phi_\text{open}/\Phi_\star$-$\Upsilon_\star$-$f$ relationship with a power-law index of $m_\Phi=-0.165$, which is again comparable to previous open flux stellar wind studies \citep[see, e.g.][]{2015ApJ...798..116R,2017ApJ...845...46F,2017ApJ...849...83P,Pantolmos2020}. Our $\beta_\Phi$ term has a proportionality constant of $K_{\Phi,2} = 0.0111$, suggesting that the open flux is more sensitive to rotation compared to the effects on wind acceleration solely (encapsulated by our $\beta_\star$ term).

Overall, for our solar-like differential rotation cases, i.e., $\omega < 1$, we find the rotation in the wind to be decreased
compared to a solid-body case with identical stellar properties (rotating at $\Omega_{\star,\text{eq}}$ with a fixed mass-loss rate), decreasing the stellar wind torque by approximately a factor of $\omega$ (ignoring centrifugal effects).\footnote{For fixed mass-loss rate, one can show $\dot{J}/\dot{J}_{\text{SB}} = (\beta_{\star,\text{SB}}/\beta_\star)^{2m_\star} \omega$, where the subscript ``SB" represents the solid-body value. For $f_\text{eq} \ll K_{\star,2}$, i.e., when centrifugal effects are negligible, $\dot{J}/\dot{J}_{\text{SB}} \approx \omega$.} We note that the parameterization of $\omega$ used in our formulation is specific to the differential rotation profile adopted; however, it is likely that one can derive $\omega$ for different profiles, and follow the framework of this paper to predict the torque.

{Differential rotation, in general, can shear small scale magnetic field structures, and affect the coronal dynamics \citep[see, e.g.,][]{Lionello_2006,Morgan_2011,Lionello_2020}. However, in this study, we solely investigate the effects of differential rotation on the rotation rate at the base of the wind, and how this affects global angular momentum loss, independent of coronal dynamics. In this case, we are concerned only about the rotation rate of open magnetic field lines (i.e., single footprints), rather than connected footprints undergoing shear. In a 3D domain, differential rotation can redistribute the open magnetic flux of a non-axisymmetric field (e.g., a tilted dipole); hence, the assumed initial magnetic flux distribution changes in time and the solution can not be stationary. On the other hand, the timescales to establish a wind solution are much shorter than the differential rotation and magnetic field reorganization timescales. Therefore, our axisymmetric results represent the solution for a wind in which the timescale of reorganization is much longer than the time to establish the (quasi-)steady-state conditions in the wind. We also note that assuming a dipolar magnetic configuration is a simplification, and real solar-like and low-mass stars exhibit smaller-scale magnetic field structures, in addition to their dipolar components \citep[e.g.,][]{doi:10.1146/annurev-astro-082708-101833}. However, various works have suggested that the open flux, hence the angular momentum loss, is dominated by the dipolar magnetic field component \citep{10.1093/mnrasl/slw206,10.1093/mnras/stw3094,10.1093/mnras/stx2599,2018ApJ...854...78F}. 
Our simulation results should apply to any real star, under the condition that the differential rotation of the open field lines is similar to the profiles we explored.
Future studies should consider more realistic magnetic field geometries and differential rotation profiles, in order to explore whether magnetic topology, closed-loop shearing, and other coronal dynamics affect the differential rotation in the wind.}

\subsection{Variations During a Solar Magnetic Cycle}\label{sec:solar_case}

We use our modified prescription to illustrate the potential effect differential rotation could have on predicting the wind torque of the Sun, using time-varying solar properties (over solar cycles 23 to 24) calculated in \citet{2018ApJ...864..125F}. Dipolar magnetic field strength is calculated using synoptic magnetograms of the surface field strength (from solar cycles 23 to 24) from two sources: the Michelson Doppler Imager at the \emph{Solar and Heliospheric Observatory (SOHO/MDI)}, and the Helioseismic and Magnetic Imager at the \emph{Solar Dynamic Observatory (SDO/HMI)}. The estimated dipolar field strength (at the pole) varies in the range $0.108-3.14$ G (average: $1.54$ G). Mass-loss rate is calculated from 27 day averages via measurements of the solar wind speed and density from the \emph{ACE} spacecraft, i.e., $\dot{M}_\odot = 4\pi \langle R^2 v_R (R) \rho(R)\rangle_{27 \, \mathrm{day}}$. The estimated mass-loss rate varies in the range $(0.697 - 4.30) \times 10^{-14}$ $M_\odot$ yr$^{-1}$ (average: $1.78 \times 10^{-14}$ $M_\odot$ yr$^{-1}$). A typical measured value of the solar equatorial sidereal rotation period is 24.47 days \citep{1990ApJ...351..309S}; this corresponds to a rotation rate of $\Omega_{\odot,\text{eq}}=2.97 \times 10^{-6}$ rad s$^{-1}$, giving $f_{\odot,\text{eq}} = 4.73 \times 10^{-3}$. We adopt a relative differential rotation rate of $\alpha_\odot = 0.3$.

\begin{figure}
\begin{center}
\includegraphics[width=0.475\textwidth]{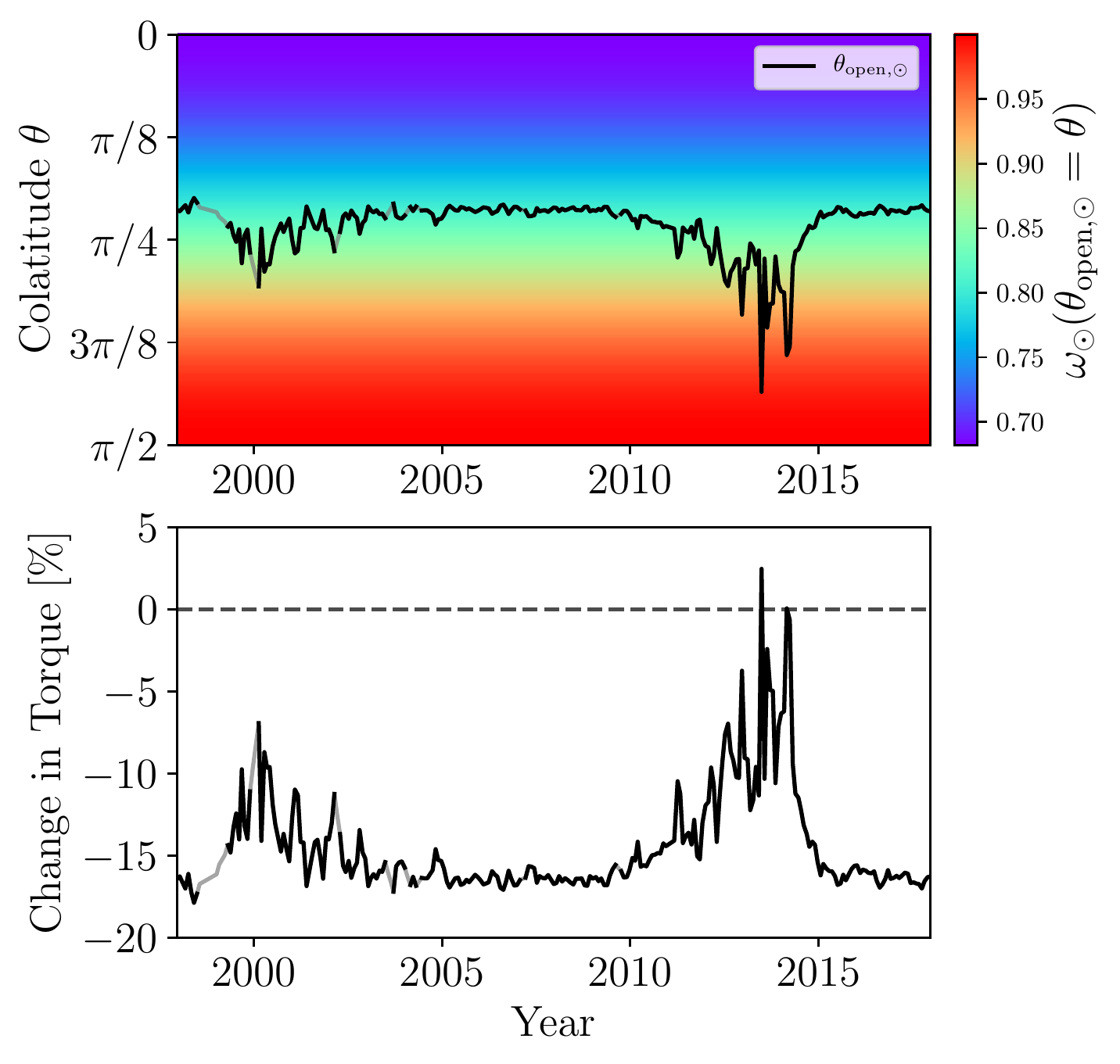}
\caption{Estimated opening colatitude of the solar wind (top), and estimated percentage change in predicted solar torque (bottom), as a function of time (in years) over solar cycles 23 and 24. \label{fig:solar}}
\end{center}
\end{figure}

Using these observations, we determine corresponding values of $\omega_\odot$ (via Equation~(\ref{eq:omega_full})) as a function of time. In the top panel of Figure~\ref{fig:solar}, we illustrate the ``effective" rotation of the wind's sensitivity to the stellar wind area, by plotting the colatitude of the opening field line (determined with $\omega_\odot$ via Equation~(\ref{eq:omega_2})) as a function of time for solar cycles 23 and 24 (black line). Grey lines are interpolated values, due to missing data for these periods. We superimpose a color map of $\omega_\odot(\theta_{\text{open},\odot}=\theta)$ (determined by substituting $\theta$ into Equation~(\ref{eq:omega_2})), which demonstrates the distribution of its theoretical value for a given colatitude of the opening field line. At the solar minima, where the stellar wind region is at its smallest (corresponding to lower opening colatitudes), $\omega_\odot\approx0.8$, i.e., the ``effective" rotation of the wind is roughly $80 \%$ of the equatorial rotation rate. At the most extreme snapshot of the solar maxima, where the stellar wind region is at its largest (corresponding to the highest opening colatitude), $\omega_\odot\approx0.99$, i.e., the ``effective" rotation of the wind is determined predominately by the most-rapid flows close to the equatorial value. 

We determine the change in the predicted solar torque using our modified prescription, compared to the solid-body prediction. 
For the solid-body approximation, we adopt the Carrington sidereal rotation period of 25.4 days, which corresponds to the period it takes for fixed features on the solar surface at a colatitude of 64$^\circ$ (typical for sunspots and other solar activity) to rotate to the same apparent position when viewed from Earth \citep[][pp. 221, 244]{carrington1863observations}; this corresponds to a rotation rate of $\Omega_{\odot,\text{SB}}=2.86\times10^{-6}$ rad s$^{-1}$, giving $f_{\odot,\text{SB}} = 4.56 \times 10^{-3}$.
Using Equation~(\ref{eq:torque_wind_full_param}), and assuming magnetocentrifugal effects on the solar wind efficiency are negligible ($\omega_\odot f_{\odot,\text{eq}} \sim f_{\odot,\text{SB}} \ll K_{\star,2}$), the ratio of our torque prediction and the solid-body prediction for the solar case can be interpreted as

\begin{align}\label{eq:solar_torque}
\frac{\dot{J}_\odot}{\dot{J}_{\odot,\text{SB}}} = \frac{\Omega_{\odot,\text{eq}}}{\Omega_{\odot,\text{SB}}} \omega_\odot.
\end{align}
In the bottom panel of Figure~\ref{fig:solar}, we plot the percentage change in the predicted solar torque as a function of time. At the solar minima, our formulation predicts a solar torque that is roughly $18 \%$ smaller than the solid-body prediction. Therefore, for the majority of the solar cycle, the open magnetic field extends over a region that is rotating slower than the solid body rotation rate, resulting in a weaker torque. However, at the most extreme snapshot of the solar maxima, our formulation actually predicts a solar torque that is roughly $2.5 \%$ larger, but this is primarily due to the fact that $\Omega_{\odot,\text{eq}} >\Omega_{\odot,\text{SB}}$. In this extreme case, where a large fraction of surface field is opened into the wind, the open region engulfs surface regions that are rotating faster than the solid body value, and thus the torque is increased.

This analysis makes use of only the axisymmetric dipole field, which may be a reasonable approximation for most of the solar cycle except for at solar maximum, where global magnetic field configurations are often non-axisymmetric and multipolar \citep{DeRosa_2012}. However, this work could be a foundation for a more detailed analysis in which the average rotation rate of the open field regions is calculated during the solar cycle, i.e., accounting for non-axisymmetry and multipolar magnetic fields. Some observations and theoretical work suggest that while the solar photosphere rotates as described, the solar corona (and thus the base of the solar wind) may in fact be rotating differently, or even potentially as a solid body \citep[see, e.g.,][]{1995SoPh..160....1I,2008ApJ...688..656G,2017_Bagashvili,Pinto_2021}. It is likely that the rotation of the solar corona determines the rotation rate of the wind, therefore it is possible that these systems are much more complex than anticipated by these formulations, and that surface differential rotation may not be noticeably responsible for any torque discrepancies between observations and theory.

\subsection{Significance of Differential Rotation for Main-Sequence Stellar Spin-Down}

Most F, G, and K MS stars are observed to have a similar or smaller relative differential rotation rate to the Sun, i.e., $\alpha < 0.3$, and most rapidly-rotating stars are observed to be rotating almost as a solid body, i.e., $\alpha \rightarrow 0$ \citep[see, e.g.,][]{2002AN....323..336C,Reiners393_2002,10.1111/j.1745-3933.2005.08587.x,Reiners_2006,2007AN....328.1030C,10.1093/mnras/stw1443}. For the cases similar to the Sun, differential rotation could affect the stellar torque on the order of tens of percent according to our prescription (as discussed in Section~\ref{sec:solar_case}). The effect of differential rotation on wind torques is expected to be even smaller for more rapidly rotating stars, as $\alpha$ is observed to decrease with rotation period. In the case of anti-solar rotation ($\alpha < 0$), which we did not explicitly explore in our parameter study, our formulation shows $\omega$ to be inversely proportional to the fractional open flux (i.e., proportional to $\Upsilon_\star/\beta_\Phi$), suggesting that MS anti-solar rotators actually spin down more efficiently. Finally, depending on the method used for determining rotation rates, it is likely that the observed rates are not generally equal to the maximum/equatorial rate, but rather some intermediate rate.  The observed rotation rates can thus be closer to the ``effective" rotation of the wind, which would reduce the error in using solid-body stellar torque predictions.

Generally, depending on the precision required, models using solid-body rotation (and indeed using observations of rotation rates where the colatitude of the rotation rate observed is not known) are probably acceptable for studying stellar spin-down, especially since other uncertainties in the modeling (e.g., knowing the mass-loss rates or magnetic field properties) are likely comparable or larger. However, for stars demonstrating relative differential rotation comparable to, or stronger than, that of the Sun, the results presented in this paper should be considered.

\acknowledgments{LGI and SPM acknowledge support from the European Research Council (ERC) under the European Union's Horizon 2020 research and innovation program (grant agreement No 682393; \textit{AWESoMeStars}: Accretion, Winds, and Evolution of Spins and Magnetism of Stars; \url{http://empslocal.ex.ac.uk/AWESoMeStars}). AJF is supported by the ERC Synergy grant ``Whole Sun", No 810218. CZ acknowledges support from the European Research Council (ERC) under the European Union's Horizon 2020 research and innovation program (grant agreement No 742095; SPIDI: Star-Planets-Inner Disk-Interactions; \url{http://spidi-eu.org}).

The authors would like to acknowledge the use of the University of Exeter High-Performance Computing (HPC) facility in carrying out this work.

We thank Andrea Mignone and others for the development and maintenance of the PLUTO code. Figures within this work are produced using the Python package Matplotlib \citep{2007CSE.....9...90H}.}

\software{Matplotlib \citep{2007CSE.....9...90H}, PLUTO \citep{0067-0049-170-1-228,2012ApJS..198....7M}.}

\bibliography{papers}

\end{document}